\RequirePackage{fix-cm}
\documentclass[smallextended]{svjour3}       
\smartqed  
\usepackage{setspace}

\usepackage{amssymb}
\usepackage{amsmath}
\usepackage{amsfonts}
\usepackage{latexsym}

\usepackage{graphicx} 
\usepackage{bmpsize}

\usepackage{caption}
\usepackage{subcaption} 
\usepackage[numbers,sort]{natbib}

\usepackage{float} 
\usepackage{hyperref} 

\journalname{Journal of Mathematical Biology}
\begin{document}

\title{A simple mechanochemical model for calcium signalling in embryonic epithelial cells}

%


\titlerunning{Calcium-Kaouri}        

\author{K. Kaouri  \and P.K. Maini \and P.A. Skourides \and N. Christodoulou  \and
    S.J. Chapman
}


\institute{K. Kaouri \at
School of Mathematics, Cardiff University, UK\\
              \email{KaouriK@cardiff.ac.uk}           
\and
          P.K. Maini\at
              Wolfson Centre for Mathematical Biology, Mathematical Institute, University of Oxford, UK
							\and
          P.A. Skourides\at Department of Biological Sciences, University of Cyprus, Cyprus
\and
         N. Christodoulou \at Department of Physiology, Development and Neuroscience, University of Cambridge, UK\\
          \and
           S.J. Chapman \at
              Oxford Centre for Industrial and Applied Mathematics, Mathematical Institute, University of Oxford, UK
}

\date{Received: date / Accepted: date}

\maketitle

\begin{abstract}
Calcium signalling is one of the most important mechanisms of information propagation in the body. In embryogenesis the interplay between calcium signalling and mechanical forces is critical to the healthy development of an embryo but poorly understood.  Several types of embryonic cells exhibit calcium-induced contractions and many experiments indicate that calcium signals and contractions are coupled via a two-way mechanochemical feedback mechanism. We present a new analysis of experimental data that supports the existence of this coupling during Apical Constriction. We then propose a simple mechanochemical model, building on early  models that couple calcium dynamics to the cell mechanics  and we replace the hypothetical bistable calcium release with modern, experimentally validated calcium dynamics. We assume that the cell is a linear, viscoelastic material and we model the calcium-induced contraction stress with a Hill function, i.e. saturating at high calcium levels. We also express, for the first time, the ``stretch-activation" calcium flux in the early mechanochemical  models as a bottom-up contribution from stretch-sensitive calcium channels on the cell membrane.   We reduce the model to three ordinary differential equations and analyse its bifurcation structure semi-analytically as two bifurcation parameters vary - the $IP_3$ concentration, and the ``strength" of stretch activation, $\lambda$. The calcium system ($\lambda=0$, no mechanics) exhibits relaxation oscillations for a certain range of $IP_3$ values. As $\lambda$ is increased the range of $IP_3$ values decreases and oscillations eventually vanish at a sufficiently high value of $\lambda$. This result agrees with experimental evidence in embryonic cells which also links the loss of calcium oscillations to embryo abnormalities. Furthermore, as $\lambda$ is increased the oscillation amplitude decreases but the frequency increases. Finally, we also identify the parameter range for oscillations as the mechanical responsiveness factor of the cytosol increases. This work addresses a very important and understudied question regarding the coupling between chemical and mechanical signalling in embryogenesis. 
\keywords {mechanochemical model\and calcium signalling\and embryogenesis \and neurulation \and
dynamical systems \and bifurcations \and relaxation oscillations \and  stretch-sensitive calcium channels}
\PACS{87.10.Ed	\and 87.10.Pq \and 87.10.Ca \and 87.10.Vg}
\subclass{MSC 34E10 \and 37G10 \and 92B05 \and 35B32}
\end{abstract}


\section{Introduction}
\label{sec:Introduction}
Calcium signalling is one of the most important mechanisms of information propagation in the body, playing an important role as a second messenger in several processes such as embryogenesis, heart function, blood clotting, muscle contraction and diseases of the muscular and nervous systems \cite{berridge2000, brini2009, dupont2016}. Through the sensing mechanisms of cells, external environmental stimuli are transformed into intracellular or intercellular calcium signals that often take the form of oscillations and waves. 


In this work we will focus on the interplay of calcium signalling and mechanical forces in embryogenesis.
During embryogenesis, cells and tissues generate physical forces, change their shape, move and proliferate \cite{lecuit2007cell}. The impact of these forces on morphogenesis is directly linked to calcium signalling \cite{hunter2014ion}. In general, how the mechanics of the cell and tissue are regulated and coupled to the cellular biochemical responseto is essential for understanding embryogenesis. 
Understanding this mechanochemical coupling, in particular when calcium signalling is involved, is also important for elucidating a wide range of other body processes, such as wound healing \cite{antunes2013coordinated, herrgen2014calcium} and cancer \cite{basson2015}.

Calcium plays a crucial role in every stage of embryonic development starting with fast calcium waves during fertilization \cite{deguchi2000spatiotemporal} to calcium waves involved in convergent extension movements during gastrulation \cite{wallingford2001calcium}, to calcium transients regulating neural tube closure \cite{christodoulou2015cell}, morphological patterning in the brain \cite{webb2007ca2+,sahu2017calcium} and apical-basal cell thinning
in the enveloping layer cells \cite{zhang2011necessary}, either in the form of calcium waves or through Wnt/$\rm Ca^{2+}$ signalling \cite{kuhl2000ca2+, kuhl2000wnt, slusarski1997interaction, slusarski1997modulation, wallingford2001calcium, herrgen2014calcium, hunter2014ion, christodoulou2015cell, Narciso2017release, suzuki2017distinct}. Crucially, pharmacologically inhibiting calcium has been shown to lead to embryo defects \cite{smedley1986calcium, wallingford2001calcium, christodoulou2015cell}. 

In many experiments actomyosin-based contractions have been documented in response to calcium release in both embryonic and cultured cells \cite{wallingford2001calcium, hunter2014ion, herrgen2014calcium, christodoulou2015cell, suzuki2017distinct}  and it has become clear that calcium is responsible for contractions in both muscle and non-muscle cells, albeit through different mechanisms \cite{cooper2000}. Cell contraction in striated muscle is mediated by the binding of $\rm Ca^{2+}$ to troponin but in non-muscle cells (and in smooth muscle cells) contraction is mediated by phosphorylation of the regulatory light chain of myosin. This phosphorylation promotes the assembly of myosin into filaments, and it increases myosin activity. Myosin light-chain kinase (MLCK), which is responsible for this phosphorylation, is itself regulated by calmodulin, a well-characterized and ubiquitously expressed protein regulated by calcium  \cite{scholey1980regulation}. Elevated cytosolic calcium promotes binding of calmodulin to MLCK, resulting in its activation, subsequent phosphorylation of the myosin regulatory light chain and then contraction. Thus, cytosolic calcium elevation is an ubiquitous signal for cell contraction which manifests in various ways \cite{cooper2000}. 

In some tissues these contractions give rise to well defined changes in cell shape. One such example is Apical Constriction (AC), an intensively studied morphogenetic process central to embryonic development in both vertebrates and invertebrates \cite{vijayraghavan2017mechanics}. In AC the apical surface of an epithelial cell constricts, leading to dramatic changes in cell shape. Such shape changes drive epithelial sheet bending and invagination and are indispensable for tissue and organ morphogenesis including gastrulation in C. elegans and Drosophila and vertebrate neural tube formation \cite{sawyer2010apical, rohrschneider2009polarity, christodoulou2015cell}.
. 

On the other hand, the ability of cells to sense and respond to forces by elevating their cytosolic calcium is well established. Mechanically stimulated calcium waves have been observed propagating through ciliated tracheal epithelial cells \cite{sanderson1981, sanderson1988, sanderson1990}, rat brain glial cells \cite{charles1991, charles1992, charles1993}, keratinocytes \cite{tsutsumi2009}, developing epithelial cells in Drosophila wing discs \cite{Narciso2017release} and many other cell types  \cite{young1999,beraheiter2005, yang2009, tsutsumi2009}. Thus, different types of mechanical stimuli, from shear stress to direct mechanical stimulation,  can elicit calcium elevation (although the sensing mechanism may differ in each case). So, since mechanical stimulation elicits calcium release and calcium elicits contractions which are sensed as mechanical stimuli by the cell,   it is clear that a two-way mechanochemical feedback between calcium and contractions should be at play. 

This two-way feedback is supported by our work here with a new analysis of data from earlier experiments conducted by two of the authors \cite{christodoulou2015cell}; we  present this analysis in detail in Section \ref{sec:Experiments}. The analysis shows that in contracting cells, in the Xenopus neural plate, calcium oscillations become more frequent and also increase in amplitude as the calcium-elicited surface area reduction progresses. This suggests that the increased tension around the contracting cell is sensed, it leads to more calcium release and in turn to more contractions, and so on. In addition, experiments in Drosophila also support the hypothesis that a mechanohemicall feedback loop is in play \cite{solon2009pulsed, saravanan2013local}. Thus, data from these two model systems clearly show that mechanical forces generated by contraction influence calcium release and the contraction cycle. The mechanosensing takes place via, as yet undefined, mechanosensory molecules which could be mechanogated ion channels, mechanosensitive proteins at adherens junctions like alpha catenin, or even integrins which have recently been shown to become activated by plasma membrane tension in the absence of ligands \cite{yao2014force,  petridou2016ligand,delmas2013mechano}.

Given the broad range of critical biological processes involving calcium signaling and its coupling to mechanical effects, modelling this mechanochemical coupling is of great interest. Therefore, we develop a simple mechanochemical model that captures the essential elements of a two-way coupling between calcium signalling and contractions in embryonic cells. The first mechanochemical models for embryogenesis were developed by Oster, Murray and collaborators in the 80s \cite{murray1984generation, oster1984mechanochemistry, murray1988, murray2001}. Calcium evolution in those early models was modelled with a hypothetical bistable reaction-diffusion process in which the application of stress can switch the calcium state from low to high stable concentration. 
We now know that the calcium dynamics are more complicated, and, so, our mechanochemical model includes instead the calcium dynamics of the experimentally verified model in \cite{atri1993single}, which captures the experimentally observed Calcium-Induced-Calcium Release process and the dynamics of the $IP_3$ receptors on the ER. this way we update the early mechanochemical models for embryonic cells in line with recent advances in calcium signalling. Note that there are many recent models of calcium signalling induced by mechanical stimulation, for example for mammalian airway epithelial cells \cite{warren2010}, for  keratinocytes  \cite{kobayashi2014}, for white blood cells \cite{yao2016}, and for retinal pigment epithelial cells \cite{vainio2015computational}. However, these models do not include a two-way coupling between calcium signalling and mechanics.

Calcium is stored and released from intracellular stores, such as the Endoplasmic Reticulum (ER), or the Sarcoplasmic Reticulum (SR), 
according to the well-established nonlinear feedback mechanism of Calcium Induced Calcium Release (CICR)\cite{dupont2016}. 
There are many models for calcium oscillations, all capturing the CICR process. Many of them model the $IP_3$ receptors on the ER in some manner, and they can be classified as Class I or Class II models \cite{dupont2016}. In all Class I models $IP_3$ is a control parameter and oscillations can be sustained at a constant $IP_3$ concentration. Oscillations exist for a window of $IP_3$ values; the oscillations are excited at a threshold $IP_3$ value and they vanish at a suffuciently high $IP_3$ value. The Atri et al. model in  \cite{atri1993single} is an established Class I model, validated with experimental findings \cite{estrada2016cellular}. (We will call this model the `Atri model' from now on.) It also has a mathematical structure that allows us to analyse our mechanochemical model semi-analytically and easily identify the parameter range sustaining calcium oscillations. Such an analysis cannot be done for other qualitatively similar, minimal Class I models as, for example, the more frequently used Li-Rinzel model \cite{liRinzel1994}; this is one of the contributions of this work.  
 
Another contribution of our work is that we interpret the ``stretch-activation" calcium flux from the outside medium, introduced in an ad hoc manner in the early mechanochemical models,  as a ``bottom-up" contribution from recently identified, stretch sensitive (stretch-activated) calcium channels (SSCCs) \cite{dupont2016, arnadottir2010eukaryotic, moore2010stretchy, hamill2006twenty}, in this way linking the channel scale with the whole cell scale. 

The paper is organised as follows. In Section \ref{sec:Experiments} we present a new analysis of experimental data which shows that calcium and contractions in embryonic cells must be involved in a two-way mechanochemical feedback mechanism. In Section \ref{sec:Mechanochemical} we develop a new mechanochemical model which captures the key ingredients of the two-way coupling.
 In Section \ref{sec:Analysis} we analyse the model. In Section \ref{sec:Model} we briefly revisit the analysis of the Atri model \cite{atri1993single} and show the bifurcation diagrams for the amplitude and frequency of calcium oscillations. In Section \ref{sec:LinStab} we perform the bifurcation analysis of the mechanochemical model, varying the $IP_3$ concentration and the strength of stretch activation, and we identify the parameter range sustaining calcium oscillations. In Section \ref{sec:HillFunction} we model the calcium-induced contraction stress with a Hill function of order $1$, and we plot the parameter range for which oscillations are sustained. In Section \ref{sec:AmplFreq} we study the amplitude and frequency of the calcium oscillations. In Section \ref{sec:mechResponse} we investigate the bifurcation diagrams as the mechanical responsiveness of the cytosol to calcium varies. In Section \ref{sec:2ndHillFun} we consider a Hill function of order $2$ and we again identify the parameter range for oscillations. 
 In Section \ref{sec:Conclusions} we summarise our conclusions and discuss further research directions.

\section{Calcium and contractions are involved in a feedback loop in Apical Constriction: a new analysis of experimental data}
\label{sec:Experiments}
There is ample experimental evidence that mechanical stimulation of cells leads to calcium elevation \cite{sanderson1981, sanderson1988, sanderson1990, charles1991, young1999,beraheiter2005, tsutsumi2009,Narciso2017release} and that, in turn, contraction of the cytosol is elicited by calcium \cite{wallingford2001calcium, hunter2014ion, herrgen2014calcium, christodoulou2015cell, suzuki2017distinct}. Calcium signaling would therefore, at least in part, be regulated by a mechanochemical feedback loop whereby calcium-elicited contractions mechanically stimulate the cell, lead to more calcium release, then to more contractions and so on. In embryogenesis, and in particular during Apical Constriction (AC), where cells contract significantly, such a feedback loop should also be at play \cite{martin2014apical}; in this work we present a new analysis of experimental data in \cite{christodoulou2015cell} which supports this. AC is a calcium-driven morphogenetic movement of epithelial tissues, central in the embryogenesis of  both vertebrates and invertebrates \cite{vijayraghavan2017mechanics}. The apical domain of epithelial cells constricts  the apical surface area, contracting, and this leads to changes in the cell geometry that drive tissue bending; in \cite{christodoulou2015cell} the formation of the neural tube in Xenopus frogs is studied and in \cite{solon2009pulsed} dorsal closure in Drosophila. 

In \cite{solon2009pulsed} the constriction of mutants that exhibit disrupted myosin activation rescues apical myosin accumulation, suggesting that mechanically stimulating the cell can elicit contractions \cite{pouille2009mechanical}. In addition, experiments using laser ablation, and other methodologies that reduce cell contractility, reveal that mechanical feedback non-autonomously
  regulates the amplitude and spatial propagation of pulsed contraction during AC \cite{saravanan2013local} and that this process is driven by calcium  \cite{pouille2009mechanical, saravanan2013local, hunter2014ion}.
Therefore, reducing contractility reduces local tension and this suppresses contraction in the control cells. This suggests that mechanical feedback is important during AC. 

Moreover, experimental evidence suggests that sensing of mechanical stimuli involves mechanogated ion channels; in Drosophila such ion channels are required for embryos to regulate force generation after laser ablation \cite{hunter2014ion}; similarly during wound healing \cite{antunes2013coordinated}. 

\begin{figure}[ht!]
\begin{center}
\includegraphics[width=0.85\linewidth]{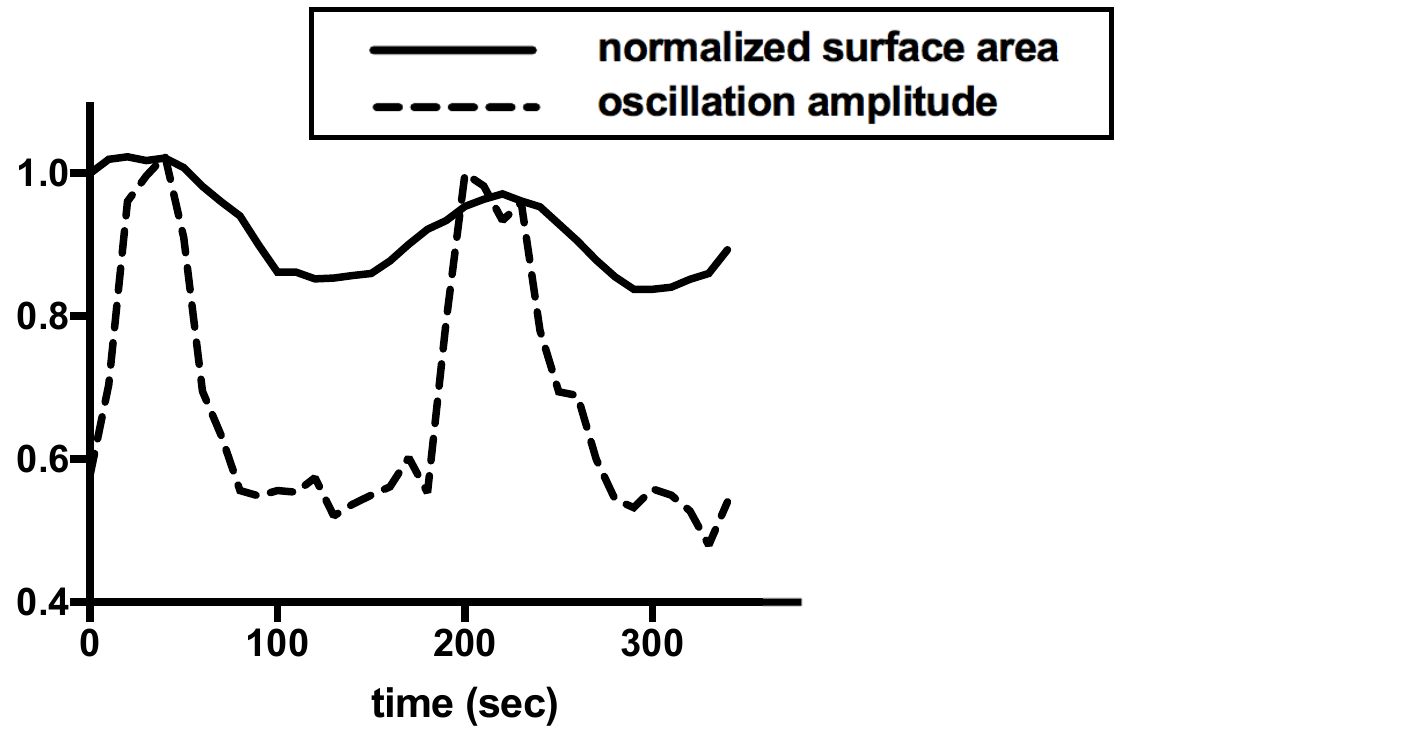}
\caption{Normalised apical surface area and amplitude of calcium oscillations in a single cell undergoing Apical Constriction. We see that calcium elevation always precedes the initiation of a contraction pulse. At $t=0$ calcium begins to rise and at $t \approx 50$ sec the surface area starts decreasing. The surface area reduction is succeeded by relaxation and stabilization of the cell at a smaller surface area. (This happens repeatedly, leading to significant reduction of the surface area over time.) See Appendix \ref{sec:suppInfo} for further details on how the Figure is produced.}
\label{fig:singlecellcontr1}
\end{center}
\end{figure}

\begin{figure}
\centering
\begin{subfigure}[t]{0.8\textwidth}
\includegraphics[width=1\linewidth]{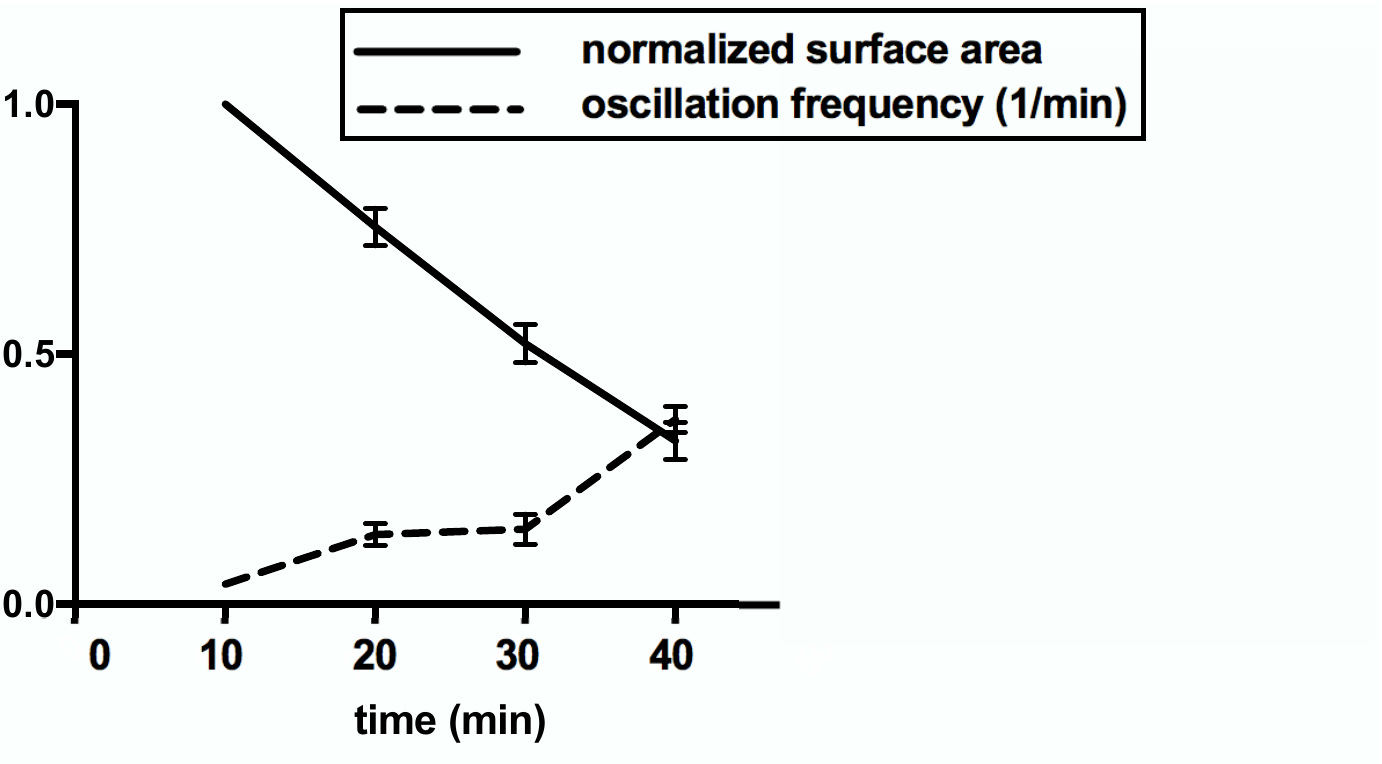}
\caption{}
\label{fig:oscillfreq_surfacearea10cells}
\end{subfigure}
\quad
\begin{subfigure}[t]{0.8\textwidth}
\includegraphics[width=1\linewidth]{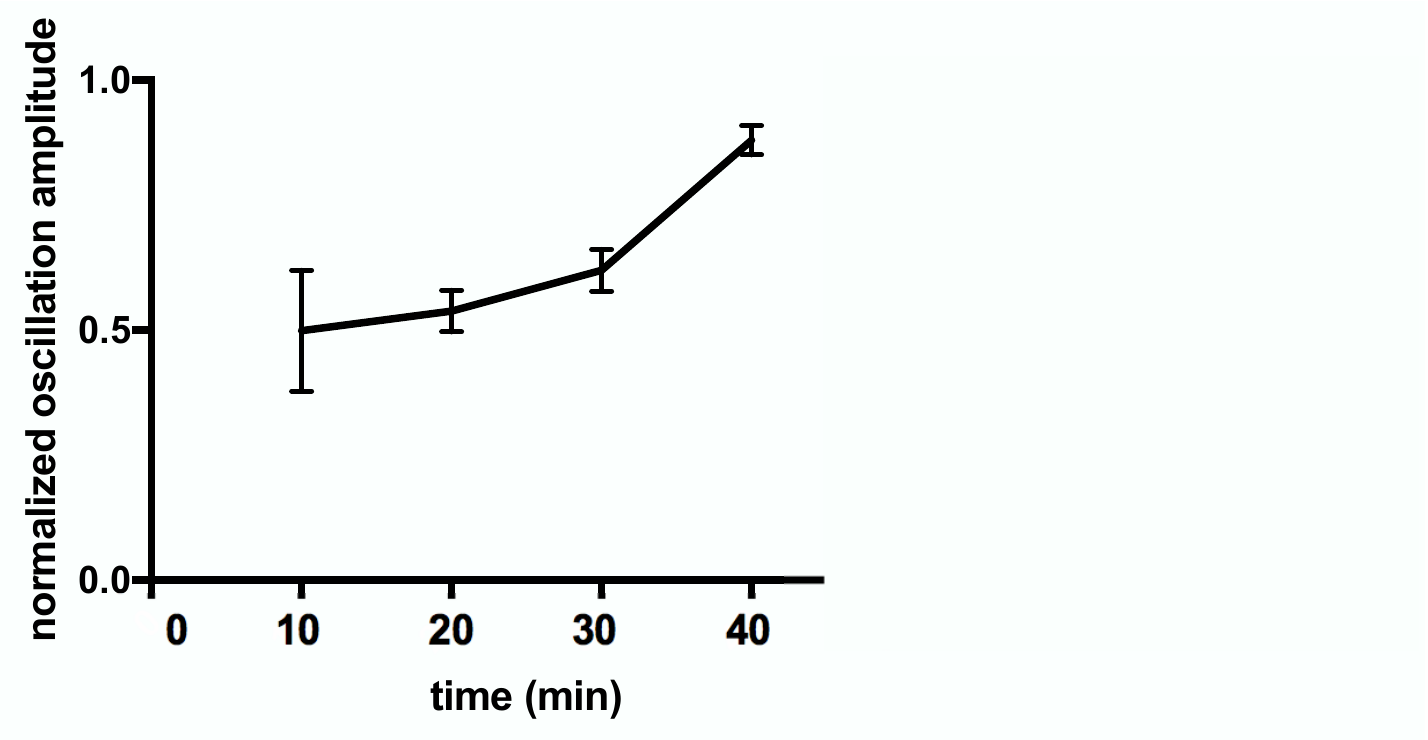}
\caption{}
\label{fig:oscillampl} 
\end{subfigure}
\caption{(a) The normalised surface area reduction is correlated to increasing oscillation frequency (10 cells).
(b) The amplitude of oscillations increases with time (10 cells). We used time lapse sequences from which the surface area of each cell was measured at $t=0$ and the average calcium oscillation frequency was calculated using a $10$-minute window (i.e. calcium oscillations for each cell were monitored between $t=0$ and $t=10$ minutes). A 10-minute window was selected so that the typical cell undergoes at least one calcium pulse. (See Appendix \ref{sec:suppInfo} for further details.)}
\label{fig:OscFreqAmplArea} 
\end{figure}

Previously, two of the co-authors have shown that cell-autonomous, asynchronous calcium transients elicit contraction pulses, leading to the pulsed reduction of the apical surface area of individual neural epithelial cells during Neural Tube Closure (NTC) in Xenopus     \cite{christodoulou2015cell}. Here, in order to investigate in detail the relationship between calcium, contraction and mechanical forces we present a new analysis of previously  collected data \cite{christodoulou2015cell}.
For a single embryonic epithelial cell (in a tissue), we plot its apical surface area and calcium level over time in Figure \ref{fig:singlecellcontr1} and we see that both oscillate, with approximately the same frequency and that the calcium pulse precedes the contraction by 30-40 seconds.  (Note that calcium oscillations emerge spontaneously without any periodic external stimulation.) More information about how Figure 1 is produced is found in Appendix \ref{sec:suppInfo}. 
In Figure \ref{fig:oscillfreq_surfacearea10cells} we plot the frequency of calcium transients and the apical surface area over time, averaged over 10 cells.  The frequency of calcium oscillations is clearly correlated with the reduction in the surface area - cells with a smaller surface area  exhibit more frequent calcium oscillations. Also,  in Figure \ref{fig:oscillampl}, for the same 10 cells and in the same timeframe, we plot the calcium oscillation amplitude, which increases with time. Therefore, the reduction in the surface area correlates also to an increase in the amplitude of the calcium oscillations. Therefore, increased surface area reduction (i.e. increased tension and hence increased mechanical stimulation) correlates with increased frequency and increased amplitude, i.e. overall increased calcium release.  

Summarising, our analysis shows that calcium oscillations trigger contraction pulses that lead to pulsed  reduction in the apical surface area over time. It also shows that the increasing localized tension in a contracting cell correlates with calcium pulses of higher frequency and larger amplitude, confirming the presence of a mechanochemical feedback loop.

\section{A new mechanochemical model for embryonic epithelial cells}
\label{sec:Mechanochemical}
We develop a simple  mechanochemical model that captures the essential components of a two-way coupling of contractions and calcium signals in embryonic epithelial cells undergoing AC.  Since the cell machinery involved in the mechanochemical coupling is similar in most cell types \cite{cooper2000} our model, with some modifications, can also be applicable to other cell types. The essential features of our model are a component modelling the cell mechanics and a component modelling calcium dynamics, coupled through a two-way feedback. Such models have been proposed by Oster, Murray and collaborators in the 80s \cite{murray1984generation, oster1984mechanochemistry, murray1988, murray2001} and here we update those models by replacing the hypothetical bistable calcium dynamics with the experimentally verified calcium dynamics in \cite{atri1993single}. We also replace the ad hoc stretch activation calcium flux in \cite{murray2001}  with a ``bottom-up" calcium release through the SSCCs, thus linking the channels' characteristics to the whole cell scale. The model takes the form
\begin{align}
\label{eq:Atri3Da}
\frac{dc}{dt}&=J_{\rm ER}-J_{\rm pump}+J_{\rm leak}+J_{\rm SSCC}\\
\label{eq:Atri3Dc}
\tau_h\frac{dh}{dt}&=\frac{k^2_2}{k^2_2+c^2}-h,\\
\label{eq:Atri3Db}
\frac{d \theta}{dt}&=-\frac{E'(1+\nu')}{(\xi_1 +\xi_2)}\theta+\frac{1}{(\xi_1 +\xi_2)}T_D(c),\\
\nonumber
\textrm{where }J_{\rm ER}&=k_f\mu(p)h\frac{bk_1+c}{k_1+c},\,\,\,J_{\rm pump}=\frac{\gamma c}{k_{\gamma}+c},\,\,\,J_{\rm leak}=\beta,\,\,\,J_{\rm SSCC}=S \theta.
\end{align}
Here, $c$ is the cytosolic calcium concentration, $h$ is the fraction of $IP_3$ receptors on the ER that have not been inactivated by calcium, and $\theta$ is the dilation/compression of the apical surface area of the cell.
In ODE \eqref{eq:Atri3Da},  $J_{\rm ER}$ models the flux of calcium from the ER into the cytosol through the $IP_3$ receptors, $\mu(p)=p/(k_{\mu}+p)$ is the fraction of $IP_3$ receptors that have bound $IP_3$ and is an increasing function of $p$, the  $IP_3$ concentration. The constant $k_f$ denotes the calcium flux when all $IP_3$ receptors are open and activated, and $b$ represents a basal current through the $IP_3$ channel.  $J_{\rm pump}$ represents the calcium flux pumped out of the cytosol where $\gamma$ is the maximum rate of pumping of calcium from the cytosol and $k_{\gamma}$ is the calcium concentration at which the rate of pumping from the cytosol is at half-maximum. $J_{\rm leak}$ models the calcium flux leaking into the cytosol from outside the cell.  Note that from now on we will neglect $J_{\rm leak}$ as this is a good approximation for the embryonic epithelial cells we consider. 

$J_{\rm SSCC}$ is the calcium flux due to the activated SSCCs. SSCCs have been identified experimentally in recent years - they are on the cell membrane and allow calcium to flow into the cytosol from the extracellular space. They are activated when exposed to mechanical stimulation and they close either by relaxation of the mechanical force or by adaptation to the mechanical force \cite{dupont2016, arnadottir2010eukaryotic, moore2010stretchy, hamill2006twenty}. The constant $S$ represents the `strength' of stretch activation. In Section \ref{sec:stretchflux} we will derive a relationship for $S$ as a function of the characteristics of a SSCC.

The inactivation of the  $IP_3$ receptors by calcium acts on a slower timescale compared to activation \cite{dupont2016} and so the time constant for the dynamics of $h$, $\tau_h>1$ in ODE \eqref{eq:Atri3Dc}. In ODE \eqref{eq:Atri3Db} $T_D(c)$ is a contraction stress that expresses how the stress in the cell depends on the cytosolic calcium level. The constants $\xi_1, \xi_2$ are, respectively, the shear and bulk viscosities of the cytosol and the constants $E'=E/(1+\nu)$ and $\nu'=\nu/(1-2\nu)$, where $E$ and $\nu$ are, respectively, the Young's modulus and the Poisson ratio.  

Our mechanochemical model is also an extension of the Atri model,
\begin{align}
\label{eq:Atria}
\frac{dc}{dt}&=J_{\rm ER}-J_{\rm pump}+J_{\rm leak},\\
\label{eq:Atrib}
\tau_h\frac{dh}{dt}&=\frac{k^2_2}{k^2_2+c^2}-h,
\end{align}
since ODE \eqref{eq:Atri3Da} is ODE \eqref{eq:Atria} but with $J_{\rm SSCC}$ added to the right hand side as an extra source term.
 The detailed derivation of the Atri model is presented in \cite{atri1993single}, where it was initially formulated, and the reader is referred there for more details. The parameter values, which we take from \cite{atri1993single}, are summarised in the Appendix, Table 1.
The Atri model  is one of the minimal Class I models, in which relaxation oscillations can be sustained at constant $IP_3$ concentration \cite{keener1998, dupont2016}. It was developed as a model for intracellular calcium oscillations in Xenopus oocytes but it has been subsequently used to model calcium dynamics in other cell types including glial cells \cite{wilkins1998}, mammalian spermatozoa \cite{olson2010}, and keratinocytes \cite{kobayashi2014, kobayashi2016}. In addition, modified Atri models have been developed in \cite{shi2008, harvey2011, liu2016}. In \cite{estrada2016cellular} the Atri model was compared to seven other calcium dynamics models and it exhibited the best agreement with experiments along with the more frequently used Li-Rinzel model \cite{liRinzel1994}. Also, the Atri model has a mathematical structure that allows us to perform a large part of our study analytically. The Atri system is also mathematically interesting because its relaxation oscillations have a different asymptotic structure to that of the well-known Van der Pol oscillator and similar excitable systems. We will present an asymptotic analysis of the Atri model and of our mechanochemical model in future work. 

Now, we describe our modelling assumptions and the remaining components of the model in more detail. 	

\subsection{Stretch-activation calcium flux}
\label{sec:stretchflux}
In the early mechanochemical models \cite{murray2001} the  stretch-activation flux $S \theta$ was introduced in a somewhat ad hoc manner. Here, we derive it in a bottom-up manner, from the contribution of the SSCCs to the cytosolic calcium concentration.

A model for the opening and closing of SSCCs was developed in \cite{vainio2015computational} for retinal pigment epithelial cells; we adapt it here for embryonic epithelial cells for which no such modelling has been performed. We denote by $C_{\rm SSCC}$ the proportion of channels in the closed state, and by $O_{\rm SSCC}$ the proportion of SSCCs in the open state. The calcium flux due to the SSCCs is proportional to the number of open channels so $J_{\rm SSCC}=K_{\rm SSCC}O_{\rm SSCC}$, where $K_{\rm SSCC}$ is the maximum calcium flux rate going through the SSCCs. As in  \cite{vainio2015computational}, we propose that the evolution of $O_{\rm SSCC}$ is governed by the ODE
\begin{align}
\label{eq:OSSCC}
\frac{d(O_{\rm SSCC})}{dt}=k_F \theta-(k_F \theta+k_B)O_{\rm SSCC},
\end{align}
where $k_F$ is the forward rate constant and $k_B$ is the backward rate constant.
We assume here that $O_{\rm SSCC}$ is quasi-steady, i.e. $O_{\rm SSCC}$ remains approximately constant as calcium rapidly evolves. This is a reasonable approximation, as discussed in Section 2.6 of \cite{dupont2016}. 
Therefore,
\begin{align}
\label{eq:OSSCC_steady}
O_{\rm SSCC}\approx \frac{k_F \theta}{k_F \theta+k_B}.
\end{align}
We linearise \eqref{eq:OSSCC_steady} since $\theta$ is small for a linear viscoelastic medium and under the additional assumption that $\frac{k_F}{k_B}$ is  of order 1 at most. We obtain
\begin{align}
\label{eq:OSSCC_steady_lin}
O_{\rm SSCC}&\approx \frac{k_F}{k_B}\theta\implies 
J_{\rm SSCC}=K_{\rm SSCC}\frac{k_F}{k_B}\theta \implies S=K_{\rm SSCC}\frac{k_F}{k_B}.
\end{align}
Therefore, we have derived, for the first time, an expression for $S$ as a combination of $K_{\rm SSCC}, k_F$ and $k_B$, linking in this way the characteristics of an SSCC to the macroscopic stretch-activation calcium flux.

\subsection{Derivation of ODE \eqref{eq:Atri3Db}}
\label{sec:derivation}
We can obtain ODE \eqref{eq:Atri3Db} from the full force balance mechanical equation for a  linear viscoelastic material. 
Linear viscoelasticity, at first glance, might not be an appropriate approximation for embryogenic tissue undergoing drastic changes but recent experiments \cite{dassow2010surprisingly} show it is reasonable.  For a Kelvin-Voigt,  linear viscoelastic material sustaining calcium-induced contractions the force balance equation can be written as follows \cite{landau1986, murray2001}:
\begin{align}
\label{eq:Atri3D-viscoelastic}
\nabla.\mathbf{\sigma}=0\Rightarrow \nabla.(\underbrace{\xi_1 \mathbf{e}_t+\xi_2 \theta_t \mathbf{I}}_\text{viscous stress}+\underbrace{E'(\mathbf{e}+\nu' \theta \mathbf{I})}_\text{elastic stress}-\underbrace{T_D(c)\mathbf{I})}_\text{contraction stress}=0,
\end{align}
where $\mathbf{\sigma}$ is the stress tensor, $\mathbf{e}=\frac{1}{2}(\nabla \mathbf{u}+\nabla \mathbf{u}^T)$ is the strain tensor, $\mathbf{u}$ the displacement vector, $\theta=\nabla.\mathbf{u}$ is the dilation/compression of the material, and $\mathbf{I}$ is the unit tensor. $T_D(c)$ is the contraction stress which depends on the cytosolic calcium \cite{scholey1980regulation}. 
In one spatial dimension $\mathbf{e}=e=\theta=\frac{\partial u}{\partial x}$ and therefore \eqref{eq:Atri3D-viscoelastic} becomes, upon integrating with respect to $x$,
\begin{align}
\label{eq:Atri3D-viscoelastic-1D}
(\xi_1 +\xi_2) \theta_t+E'(1+\nu') \theta -T_D(c))=A.
\end{align}
The constant of integration $A=0$ since when $c=0$, $T_D=0$ , $\theta=0$ and $\theta_t=0$. Hence, we obtain ODE \eqref{eq:Atri3Db}.

\subsection{Nondimensionalised model}
We nondimensionalise the mechanochemical model using $c=k_1\bar c$ and $t=\tau_h \bar t$. Dropping bars for notational convenience we obtain
\begin{align}
\label{eq:Atri3Da-ND}
\frac{dc}{dt}&=\mu hK_1\frac{b+c}{1+c}-\frac{\Gamma c}{K+c}+\lambda \theta=R_1(c,\theta,h;\mu,\lambda),\\
\label{eq:Atri3Dc-ND}
\frac{dh}{dt}&=\frac{K^2_2}{K^2_2+c^2}-h=R_3(c,h),\\
\label{eq:Atri3Db-ND}
\frac{d \theta}{dt}&=-k_{\theta}\theta+\hat T(c)=R_2(c,\theta),
\end{align}
In \eqref{eq:Atri3Da-ND} $K_1=k_f\tau_h/k_1$, $\Gamma=\gamma \tau_h/k_1$, $K=k_{\gamma}/k_1$, and $\lambda=\tau_h S/k_1$. In \eqref{eq:Atri3Db-ND}, $\displaystyle{k_{\theta}=\frac{\tau_h E'(1+\nu')}{(\xi_1 +\xi_2)}}$ and $T(c)=\frac{\tau_h}{(\xi_1 +\xi_2)}T_D(c)$, and in \eqref{eq:Atri3Dc-ND} $K_2=k_2/k_1$.
Using the parameter values of \cite{atri1993single} (see Appendix, Table \ref{tab:Table1}), we obtain $K_2=1$,  $\Gamma=40/7=5.71$, and $K=1/7$. Also, taking values of $E$, $\nu$ and of the viscosity from \cite{zhou2009actomyosin} ($E=8.5$ Pa, $\nu=0.4$ and $\xi_1+\xi_2=100$ Pa.s we find that $k_{\theta}$ is between 0.36 and 0.75.  For simplicity, and since the parameter values for the calcium dynamics are anyway approximate, we fix $k_{\theta}=1$.  Furthermore, $\displaystyle{T(c)=\frac{\tau_h}{(\xi_1 +\xi_2)}T_D(c)
=\frac{\tau_h}{(\xi_1 +\xi_2)}T_{0D}\hat T(c)}$ where  $\hat T(c)$ is nondimensional, and we also fix $\frac{\tau_h}{(\xi_1 +\xi_2)}T_{0D}=1$.  To our knowledge, there are no measured properties for SSCCs and therefore we take the `strength' of stretch activation as a bifurcation parameter, to explore the behaviour of the model for a range of values. 


\section{Analysis of the model}
\label{sec:Analysis}

\subsection{The bifurcation diagrams of the Atri model (no mechanics)}
\label{sec:Model}
The nondimensional Atri system is
\begin{align}
\label{eq:Atri1}
\frac{dc}{dt}&=\mu hK_1\frac{b+c}{1+c}-\frac{\Gamma c}{K+c}=F(c,h),\\
\label{eq:Atri2}
\frac{dh}{dt}&=\frac{K_2^2}{K_2^2+c^2}-h=G(c,h).
\end{align}
In Appendix \ref{sec:LinStabDetails} we carry out a linear stability analysis of \eqref{eq:Atri1}--\eqref{eq:Atri2} and a bifurcation analysis with $\mu$ as the bifurcation parameter and we find that the parameter range for relaxation oscillations (limit cycles) is $0.289 \leq \mu \leq 0.495$, as in \cite{atri1993single}.  In Appendix \ref{sec:LinStabDetails} more details on the bifurcation structure of the system are given. 

\begin{figure}
\centering
\begin{subfigure}[t]{1\textwidth}
\includegraphics[width=1\linewidth]{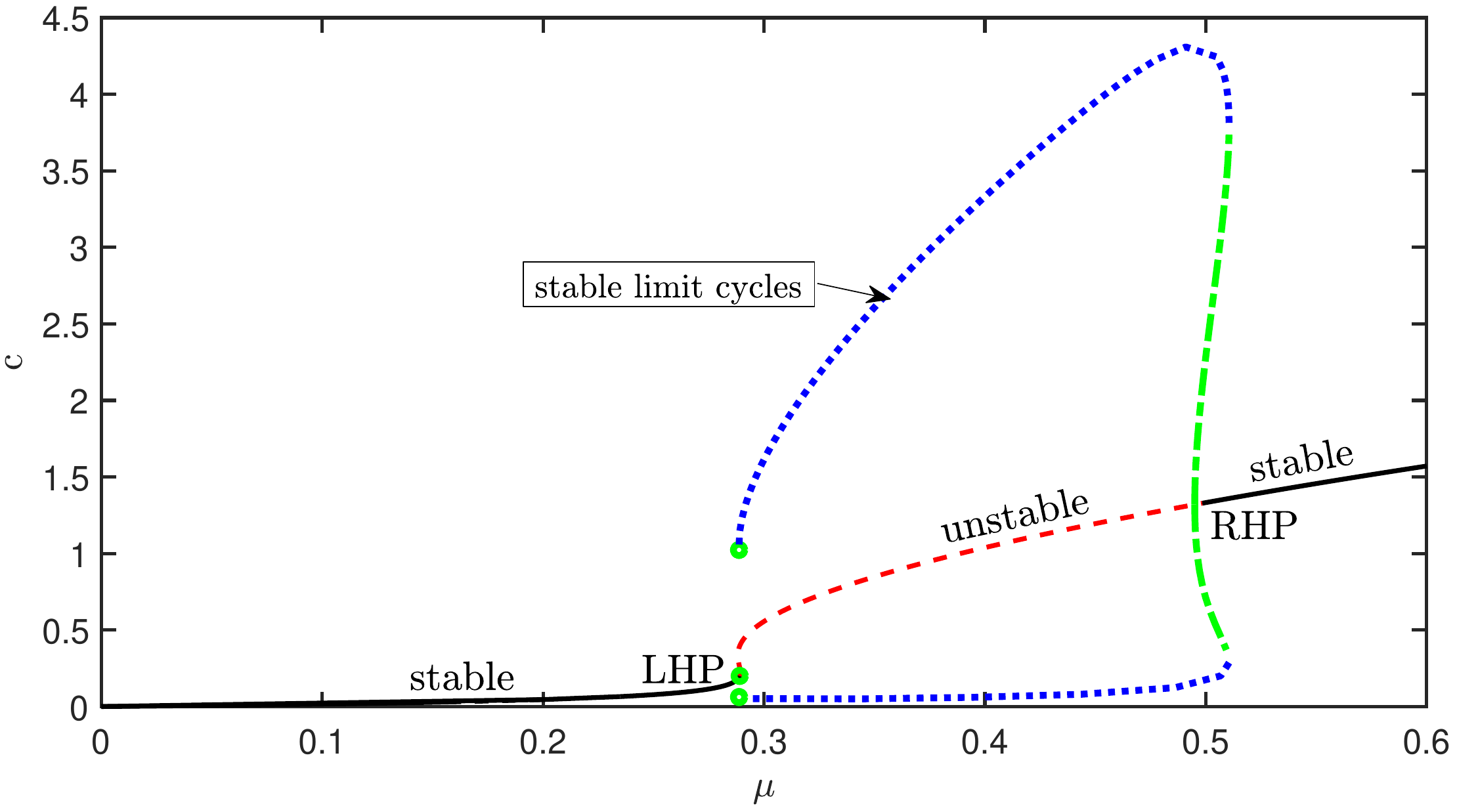}
\caption{}
\label{fig:Atri}
\end{subfigure}
\quad
\begin{subfigure}[t]{1\textwidth}
\includegraphics[width=1\linewidth]{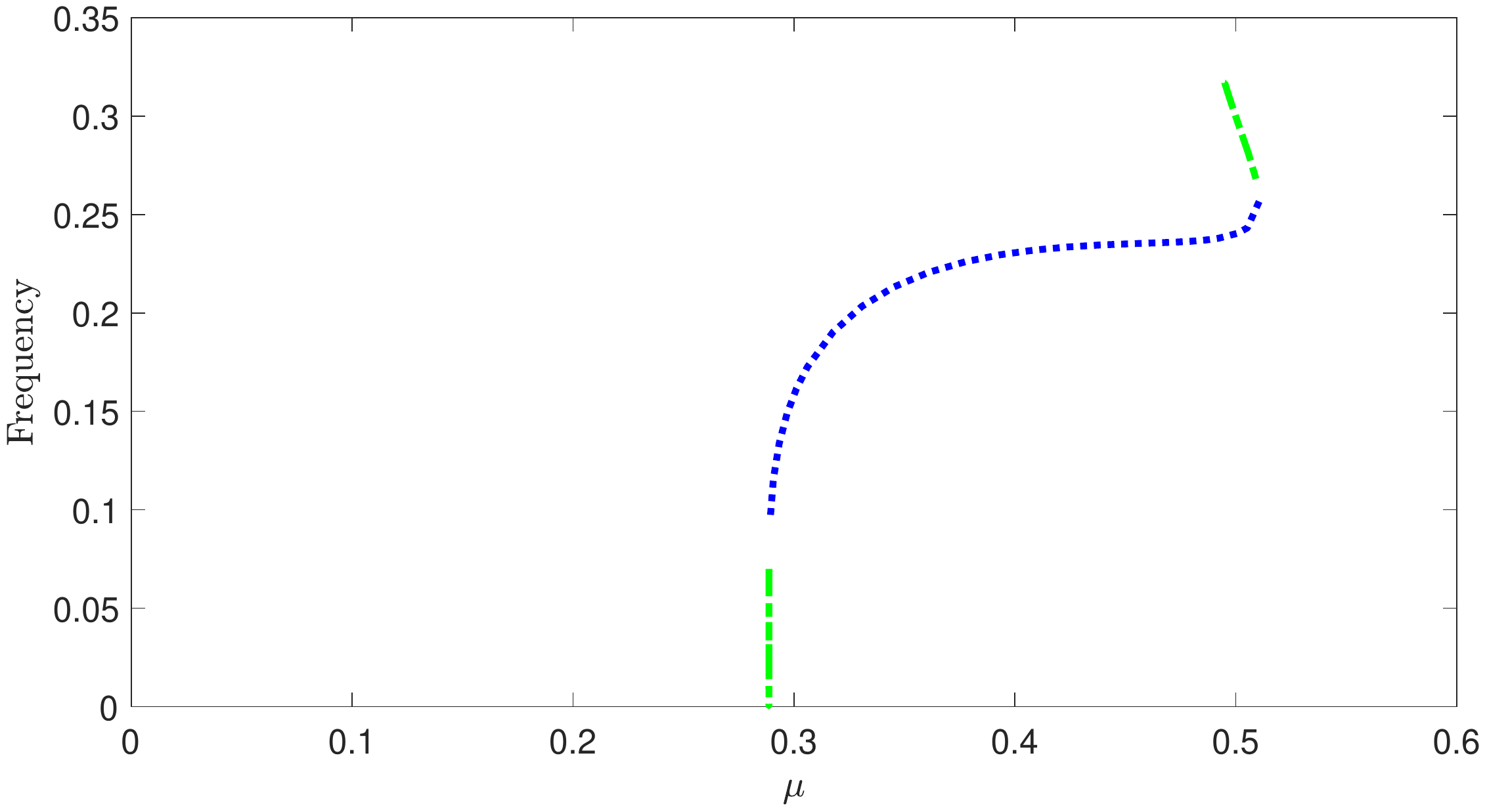}
\caption{}
\label{fig:AtriFrequency} 
\end{subfigure}
\caption{Bifurcation diagrams for the ODEs \eqref{eq:Atri1}--\eqref{eq:Atri2}, as $\mu$ ($IP_3$ level) increases: (a) Amplitude of calcium oscillations (limit cycles). The dots represent stable limit cycles and the dash-dotted part corresponds to unstable limit cycles (respectively blue and green colour online). The left Hopf point (LHP) and the right Hopf point (RHP) are indicated. 
(b) Frequency of the limit cycles.}
\label{fig:AtriBifnFrequency} 
\end{figure}

In Figure \ref{fig:AtriBifnFrequency}  we plot the bifurcation diagrams for the Atri system. In Figure \ref{fig:Atri} we present the amplitude of oscillations. The left Hopf point (LHP) and the right Hopf point (RHP) are, respectively, at $\mu=0.289$ and $\mu=0.495$. There are stable limit cycles and unstable limit cycles. The amplitude of oscillations increases with $\mu$ except for a small range of $\mu$ values near the RHP. In Figure \ref{fig:AtriFrequency} the frequency of the stable and of the  unstable limit cycles are shown, respectively. The range of $\mu$  for which both a stable and an unstable limit cycle are sustained is clearly visible as the double-valued part of the curve. The limit point of oscillations at $\mu=0.511$, where the stable and unstable limit cycle branches coalesce, is also visible. The frequency of the stable limit cycles increases \emph{slowly} as $\mu$ increases and the lower, stable branch approximates the square root of $\mu$, as predicted by bifurcation theory \cite{kuznetsov2013}. 
\subsection{Linear stability analysis of the mechanochemical model}
\label{sec:LinStab}
The steady states of the system \eqref{eq:Atri3Da-ND}--\eqref{eq:Atri3Dc-ND} satisfy
\begin{align}
\label{eq:mu-SS-3D}
\mu K_1\frac{1}{1+c^2}\frac{b+c}{1+c}-\frac{\Gamma c}{K+c}+\lambda \hat T(c)=0.
\end{align}
For any $\hat T(c)$, using \eqref{eq:mu-SS-3D}, we can easily plot the steady states as a function of $\mu$ and $\lambda$. The Jacobian of \eqref{eq:Atri3Da-ND}--\eqref{eq:Atri3Dc-ND} is given by
\begin{align}
  M_1=\begin{bmatrix}
    R_{1c} & \lambda& R_{1h}  \\
      \hat T'(c)& -1 & 0 \\
   R_{3c}  & 0& -1
  \end{bmatrix},
\end{align}
and the characteristic polynomial is conveniently factorised as
\begin{align}
\label{eq:factorisedpoly}
(1+\omega)(\lambda \hat T'(c)+(R_{1c}-\omega)(1+\omega)+R_{1h}R_{3c})=0,
\end{align}
where $\omega$ represents the eigenvalues. 
As one eigenvalue is always equal to -1, the bifurcations of the system can be studied through the \emph{quadratic}
\begin{align}
\omega^2-\omega(R_{1c}-1)-R_{1c}-R_{1h}R_{3c}-\lambda T'(c)=0.
\end{align}
To identify the $\mu$-$\lambda$ parameter range sustaining oscillations we seek the Hopf bifurcations which satisfy Tr$(M_2)=0$, Det$(M_2)>0$, Discr$(M_2)<0$, where 
\begin{align}
  M_2=\begin{bmatrix}
    R_{1c} & \lambda  \\
      \hat T'(c)& -1  
  \end{bmatrix}.
\end{align}
\begin{align}
\label{eq:MUvsC}
\textrm{Setting Tr}(M_2)=0 \implies \mu(c)&=\frac{(1+c^2)(1+c)^2}{K_1(1-b)}\left(1+\frac{\Gamma K}{(K+c)^2}\right),
\end{align}
and substituting in \eqref{eq:mu-SS-3D} we obtain
\begin{align}
\label{eq:LAMBDAvsC}
\lambda(c)=\frac{1}{\hat T(c)}\left(\frac{\Gamma c}{K+c}-\frac{(b+c)(1+c)}{1-b}\left(1+\frac{\Gamma K}{(K+c)^2}\right)\right).
\end{align}
Hence, we can easily  obtain the \emph{Hopf curve}, for \emph{any} $\hat T(c)$ by parametrically plotting \eqref{eq:MUvsC} and \eqref{eq:LAMBDAvsC}, with $c$ as a parameter. The interior of the Hopf curve corresponds to an unstable spiral and approximates the $\mu$-$\lambda$ parameter space sustaining oscillations (limit cycles) in the full nonlinear system. 

We also determine parametric expressions for the \emph{fold curve}. Inside the fold curve there are three steady states, on the fold curve two of states coalesce, and outside the fold curve there is one steady state. Setting Det$(M_2)=0$
\begin{align}
\label{eq:Det1}
& \implies \mu\frac{K_1}{(1+c)(1+c^2)}\left(\frac{1-b}{1+c}-2c(b+c) \right)+\lambda T'(c)=\frac{\Gamma K}{(K+c)^2}.
\end{align}
Equations \eqref{eq:Det1} and \eqref{eq:mu-SS-3D} constitute a linear system for $\mu$ and $\lambda$, so we again easily derive parametric expressions for $\mu(c)$ and $\lambda(c)$.  

Similarly, to determine parametric expressions for the curve on which Discr$(M_2)$ changes sign we set Discr$(M_2)=0$
\begin{align}
\label{eq:Discr}
\implies (R_{1c}+1)^2+4R_{1h}R_{3c}+\lambda \hat T'(c)=0,
\end{align}
which is quadratic in $\mu$ and linear in $\lambda$. Combining \eqref{eq:Discr} with \eqref{eq:mu-SS-3D} we can again determine parametric expressions for $\mu$ and $\lambda$.
In summary, we have developed a quick method for determining the three key curves characterising the geometry of the bifurcation diagram, for \emph{any} $\hat T(c)$. 

It is of course, a fortunate accident of construction that we can obtain these analytical expressions for this particular model. Since our model is qualitatively similar to any other mechanochemical model that is based on Class I calcium models, the analytical progress we make here is very useful since the insights gained from it can be applied to other mechanochemical models. A different model would have a more complex set of linear stability equations, that look quite different, but that are fundamentally saying the exact same thing. Crucial to the behaviour is the shape of the manifolds rather than the detail of the algebraic expressions.
\section{Illustrative examples}
\subsection{Contraction stress is a Hill function $\hat T(c)$ of order 1}
\label{sec:HillFunction}
\subsubsection{Hopf curves}
We assume that the calcium-induced stress $\hat T(c)$ is the Hill function
\begin{align}
\label{eq:Tmodel}
\hat T(c)=\frac{\alpha c}{1+\alpha c},\,\,\,\,\,\,\,\,\alpha>0,
\end{align}
assuming that when the calcium level increases sufficiently the stress saturates to a constant value, $T_s=1$. This is a reasonable assumption since the cells reach a point at which they stop responding mechanically to calcium since the molecules involved  in contraction, calmodulin and myosin light chain kinases, saturate for sufficiently high calcium levels \cite{stefan2008allosteric}. Also, $\hat T=0$ when $c=0$, i.e. we assume no contraction stress without calcium. $\hat T'(0)=\alpha$ is the rate of increase of $\hat T$ at $c=0$ and $1/\alpha$ is the scale of `ascent' to the saturation level $T_s$.  Therefore, we can call $\alpha$ the `mechanical responsiveness factor' of the cytoskeleton to calcium.

\begin{figure}
\begin{center}
\includegraphics[width=1\textwidth]{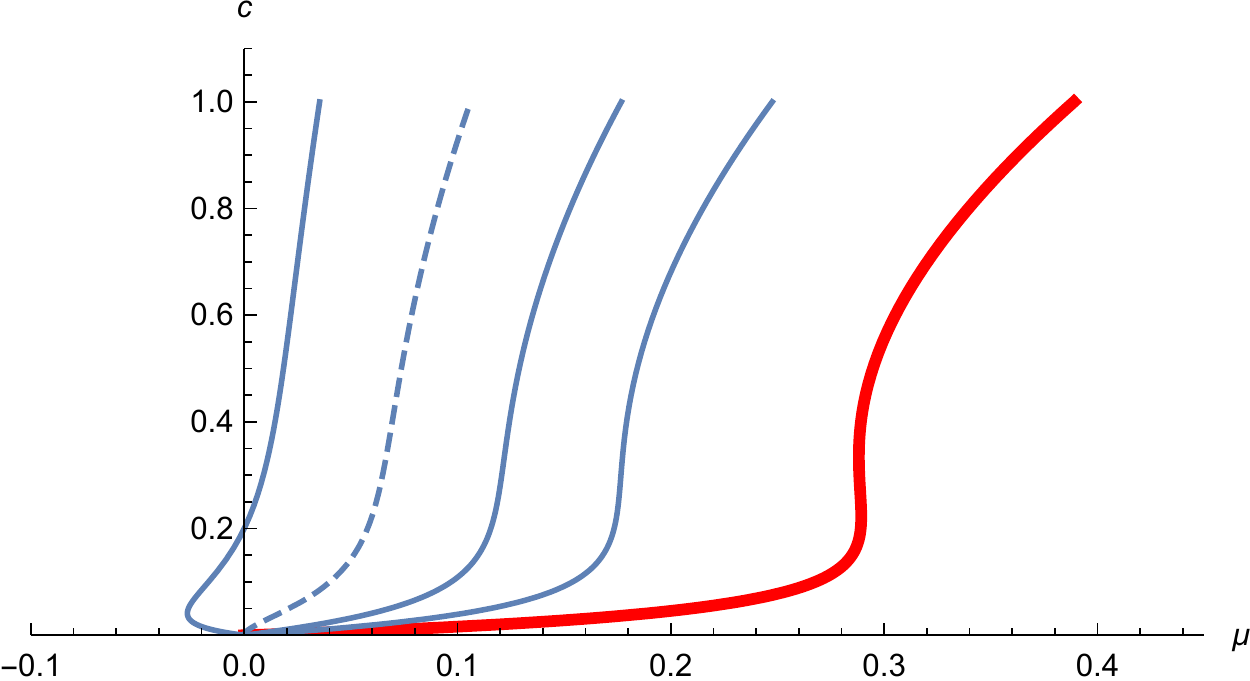}
\caption{Steady states of the system \eqref{eq:Atri3Da-ND}--\eqref{eq:Atri3Dc-ND} when $T=10c/(1+10c)$  as $\mu$ is increased, for selected $\lambda=0, 2, 3, 4, 5$--from right to left, the thick (red) curve is for
$\lambda=0$ and the dashed curve is for $\lambda=4$. (Plot done with Mathematica.)}
\label{fig:pBifDiagfmuLam}
\end{center}
\end{figure}

Choosing  $\displaystyle{\hat T(c)=10c/(1+10c)}$ as an illustrative example, in Figure \ref{fig:pBifDiagfmuLam} we use \eqref{eq:mu-SS-3D} to plot the steady state as a function of $\mu$, for selected  increasing values of $\lambda$ (equilibrium curve). For $\lambda<4$ the equlibrium curve is qualitatively similar to that of the Atri model (see Figure \ref{fig:Atri}) but at $\lambda=4$ the curve changes qualitatively and a second non-zero steady state exists for $4<\lambda<40/7$, and a part of the curve corresponds to negative values of $\mu$ (see Appendix \ref{sec:muZero} for details). For $\lambda > 40/7$ no steady state exists for positive $\mu$ and hence $\lambda \le 40/7$ is the biologically relevant range of the model, for $\alpha=10$ (see Appendix \ref{sec:muZero}). 

\begin{figure}[t]
		\centering
		\includegraphics[width=0.85\linewidth]{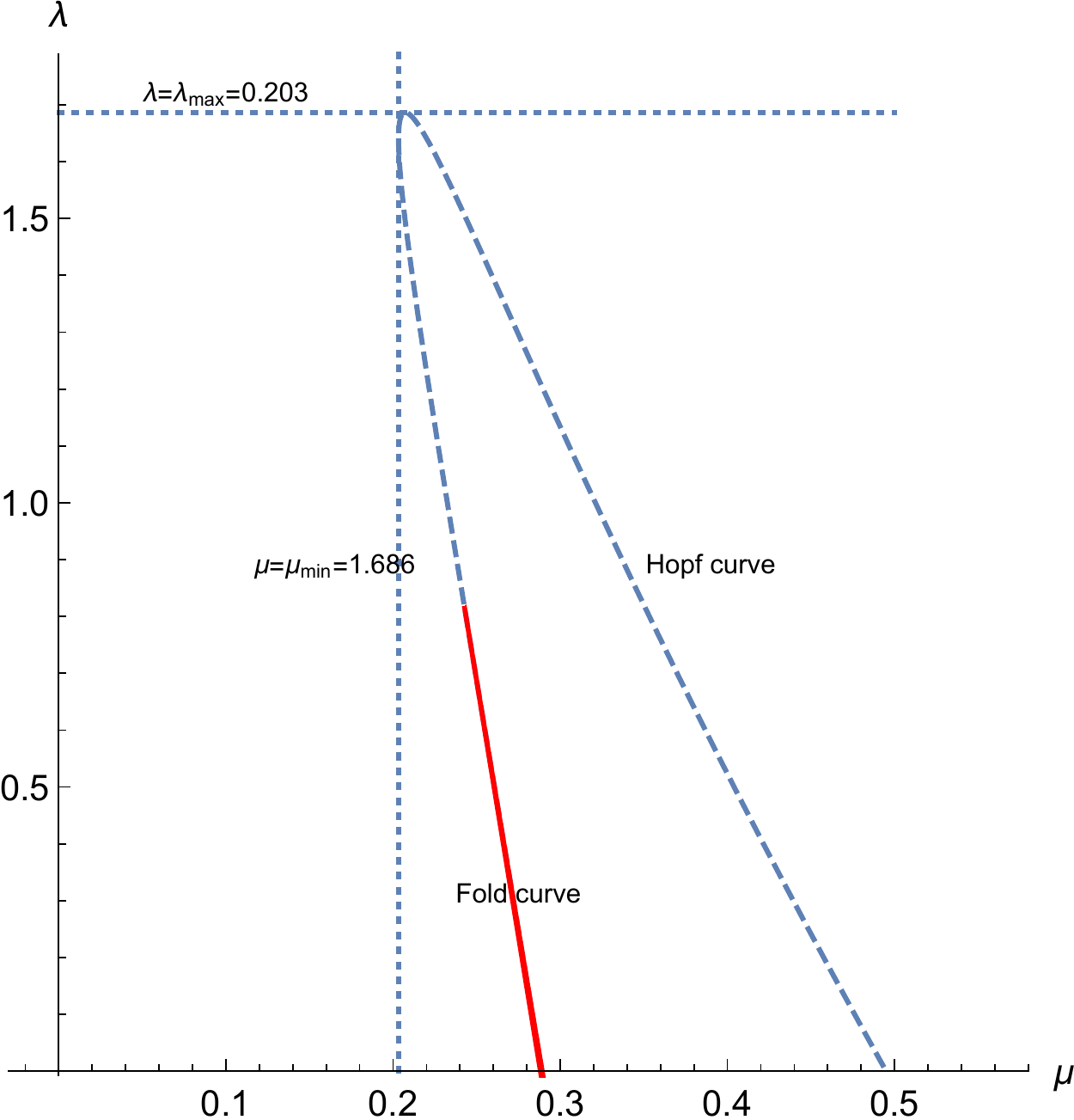}
		\caption{Geometry of the bifurcation diagrams of the system \eqref{eq:Atri3Da-ND}--\eqref{eq:Atri3Dc-ND} when $\hat T(c)=10c/(1+10c)$, plotted using the analytical,  parametric expressions we derived for  $\mu$ and $\lambda$. The Hopf curve is dashed (blue colour online) and the  fold curve is drawn with a solid line (red colour online). The horizontal and vertical dashed lines correspond, respectively, to the maximum value of $\lambda$, $\lambda_{\rm \max}=1.686$ and to the minimum value of $\mu$, $\mu_{\rm min}=0.203$ for which oscillations can be sustained. (Plot done with Mathematica.)}\label{fig:HopfCurveFold}		
\end{figure}

In Figure \ref{fig:HopfCurveFold} we plot the Hopf curve and the fold curve.  We observe the following: (i) for $\lambda=0$ we recover the Hopf points and the fold points of the Atri model, as expected. (ii) As $\lambda$ increases the range of $\mu$ that sustains oscillations decreases. There is a global minimum value of $\mu$ that can sustain oscillations, $\mu_{\rm min}$. 
(iii) \emph{The oscillations are suppressed} for a critical maximum value of $\lambda$, $\lambda_{\rm \max}$ and the system is in a high calcium state. 
Overall, we conclude the following from the Hopf curve:
\begin{itemize}
\item for low $IP_3$ values the Atri system does not sustain oscillations but there are two possibilities for the mechanochemical model as $\lambda$ increases: \\ 
\noindent$\bullet$ for $\mu<\mu_{\rm min}$ no increase in $\lambda$ will ever elicit oscillations. \\ 
\noindent$\bullet$ for $0.203=\mu_{\rm min}<\mu < 0.289$ when $\lambda$ reaches a certain value, $\lambda_{\rm OSC}$, oscillations are elicited, and $\lambda_{\rm OSC}$ decreases as $\mu$ approaches $0.289$. The oscillations vanish at a larger value of $\lambda$. 
\item for $IP_3$ values for which the Atri system sustains oscillations ($0.289<\mu<0.495$)  in the mechanochemical model oscillations eventually vanish at a critical 
$\lambda$. This critical $\lambda$ decreases monotonically as $\mu$ increases towards $0.495$. 
\item for high $IP_3$ values ($\mu \geq 0.495$) no oscillations are sustained in the Atri system and a further increase in $\lambda$ will never elicit oscillations. 
\end{itemize}
Therefore, for fixed cytoskeletal mechanical responsiveness factor, $\alpha=10$, and for fixed parameter values as in \cite{atri1993single} a range of behaviours emerge as $\mu$ and $\lambda$ vary:   at low $IP_3$ levels that do not elicit oscillations in the Atri system mechanical effects can elicit oscillations, for intermediate $IP_3$ levels that do sustain oscillations in the Atri system increasing mechanical effects always leads to the oscillations vanishing, and for high $IP_3$ levels that cannot sustain oscillations in the Atri system no amount of stretch activation can ever elicit oscillations.

Overall, we conclude that in this case the effect of mechanics  can significantly affect calcium signalling. A very important prediction of the model is that oscillations vanish for sufficiently large stretch activation.  This prediction 
agrees with the experiments reported in \cite{christodoulou2015cell} (Figure 5D); when the cells were forced to enter a high, non-oscillatory calcium state they monotonically reduced their apical surface area without oscillations. Interestingly, although the loss of oscillations did not affect the reduction of the apical surface on average, it led to  the disruption of the patterning involved in AC and neural tube closure failed, leading to severe embryo abnormality. 

In fact, the model also agrees, qualitatively, with other experimental observations. Intracellular calcium levels (which are regulated by $IP_3$) directly affect cell contractility \cite{christodoulou2015cell}. At low levels of $IP_3$ and hence low levels of calcium, cells are not able to contract and therefore apical constriction does not take place. At a threshold $IP_3$ value the system changes behaviour and calcium oscillations/transients appear (mathematically this corresponds to a \emph{bifurcation}). The calcium oscillations enable the ratchet-like pulsating process of the AC to progress normally.  At high levels of $IP_3$ the cell has been shown to enter a high-calcium state with no oscillations, as mentioned above. (This corresponds to another bifurcation since the system changes its qualitative behaviour.) 

Regarding bistability, note that the fold curve consists of two branches very close to each other since the Atri system is bistable for a very small range of $IP_3$ concentrations. As $\lambda$ increases this range decreases and eventually vanishes at $\lambda \approx 0.83$, where the two fold curve branches merge.

Summarising, the parametric method we have developed allows us to easily plot the Hopf curve, and the two other important curves of the bifurcation diagram, for any functional form of $\hat T(c)$ we may choose, and thus examine quickly the effect of mechanics on calcium oscillations. We note that in the experiments of \cite{christodoulou2015cell} the calcium-induced stress saturates to a non-zero level as calcium levels increase and hence we chose a $\hat T(c)$ that saturates.  In other cell types it is possible that the cell can relax back to baseline stress and in such a case $\hat T(c)$ would not be described by a Hill function, and more experiments should be undertaken to determine the appropriate form of $\hat T(c)$.

\subsubsection{Amplitude and frequency of the calcium oscillations}
\label{sec:AmplFreq}
We now determine numerically the amplitude and frequency of oscillations (limit cycles) of the system \eqref{eq:Atri3Da-ND}--\eqref{eq:Atri3Dc-ND} when $\hat T(c)=10c/(1+10c)$. 

\begin{figure}	
	\centering
	\begin{subfigure}[t]{1\textwidth}
		\centering
		\includegraphics[width=1\linewidth]{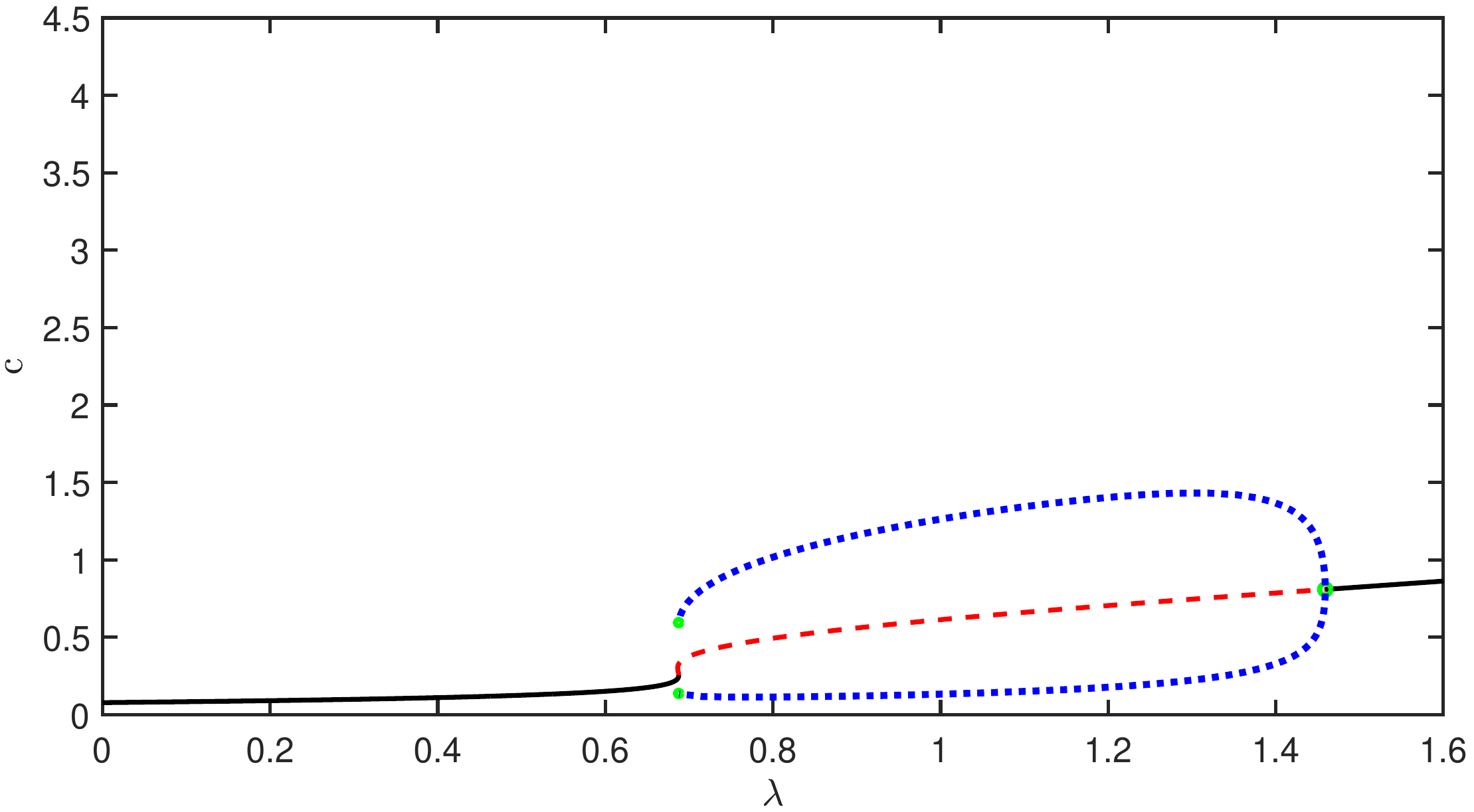}
		\caption{$\mu=0.25$}\label{fig:AtriMechsMu0Pt25}		
	\end{subfigure}
	\quad
	\begin{subfigure}[t]{1\textwidth}
		\centering
		\includegraphics[width=1\linewidth]{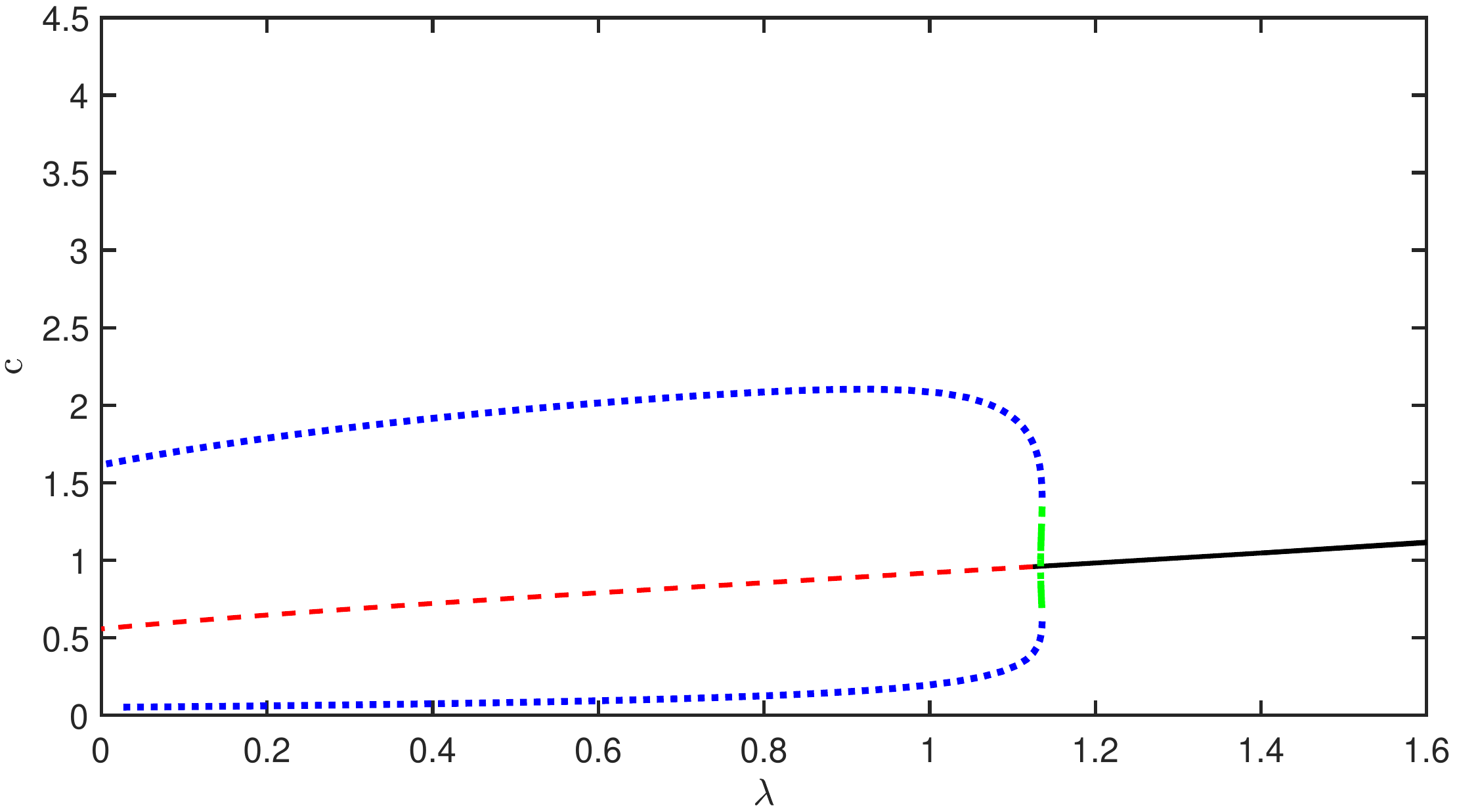}
		\caption{$\mu=0.3$}\label{fig:AtriMechsMu0Pt3}
	\end{subfigure}
	\caption{Amplitude of calcium oscillations for the system \eqref{eq:Atri3Da-ND}--\eqref{eq:Atri3Dc-ND} when  $\hat T(c)=\frac{10 c}{1+10 c}$, as $\lambda$ is increased, for: (a) $\mu=0.25$ (b) $\mu=0.3$.  The stable limit cycles are represented by dots and the unstable limit cycles by the dash-dotted parts (respectively with blue and green colour online). The plots are computed with XPPAUT and exported to Matlab for plotting.}
\label{fig:AtriMechsMu}
\end{figure}

In Figure \ref{fig:AtriMechsMu} we plot the oscillation amplitude as a function of $\lambda$, for two selected values of $\mu$, using XPPAUT. For $\mu=0.25$ (Figure \ref{fig:AtriMechsMu0Pt25}) the Atri system has no oscillations but stable limit cycles arise in the mechanochemical model  as $\lambda$ is increased, which agrees with the Hopf curve in Figure \ref{fig:HopfCurveFold}. For $\mu=0.3$ (Figure \ref{fig:AtriMechsMu0Pt3}) the Atri system has a stable limit cycle and as $\lambda$ increases, stable and unstable limit cycles emerge for a finite $\lambda$-interval, and oscillations eventually vanish for sufficiently large $\lambda$. For $\mu=0.4$ the Atri system has a stable limit cycle and as $\lambda$ increases, stable and unstable limit cycles emerge for a finite range of $\lambda$ values, and oscillations eventually vanish for a large enough value of $\lambda$. 

The oscillation amplitude changes slowly with $\lambda$ for a fixed $\mu$, that is the oscillation amplitude is robust to changes in stretch activation. 

\begin{figure}	
	\centering
     \begin{subfigure}[t]{1\textwidth}
		\centering
		\includegraphics[width=1\linewidth]{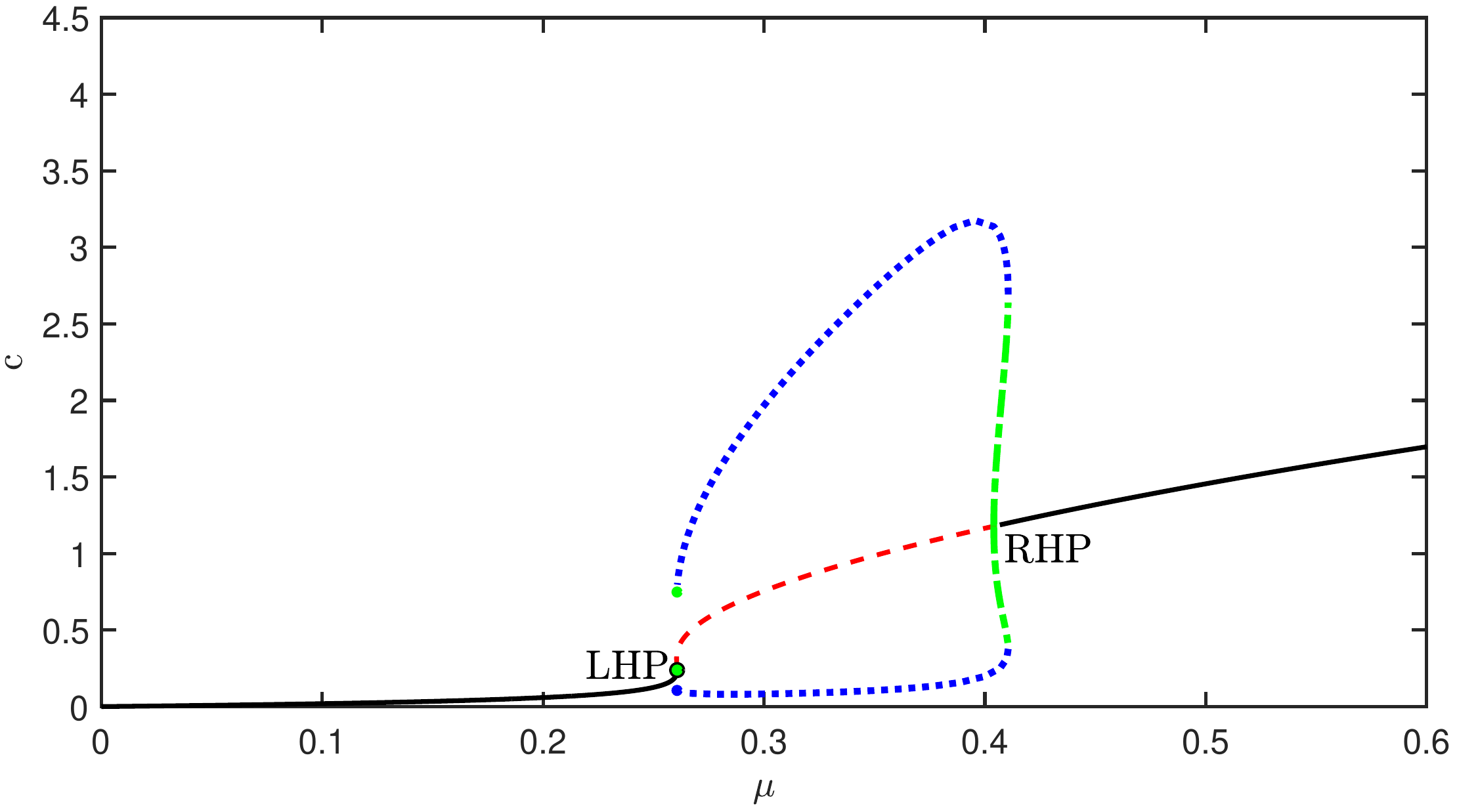}
		\caption{$\lambda=0.5$}\label{fig:AtriMechsLam0Pt5} 	
	\end{subfigure}
	\quad	
\begin{subfigure}[t]{1\textwidth}
		\centering
		\includegraphics[width=1\linewidth]{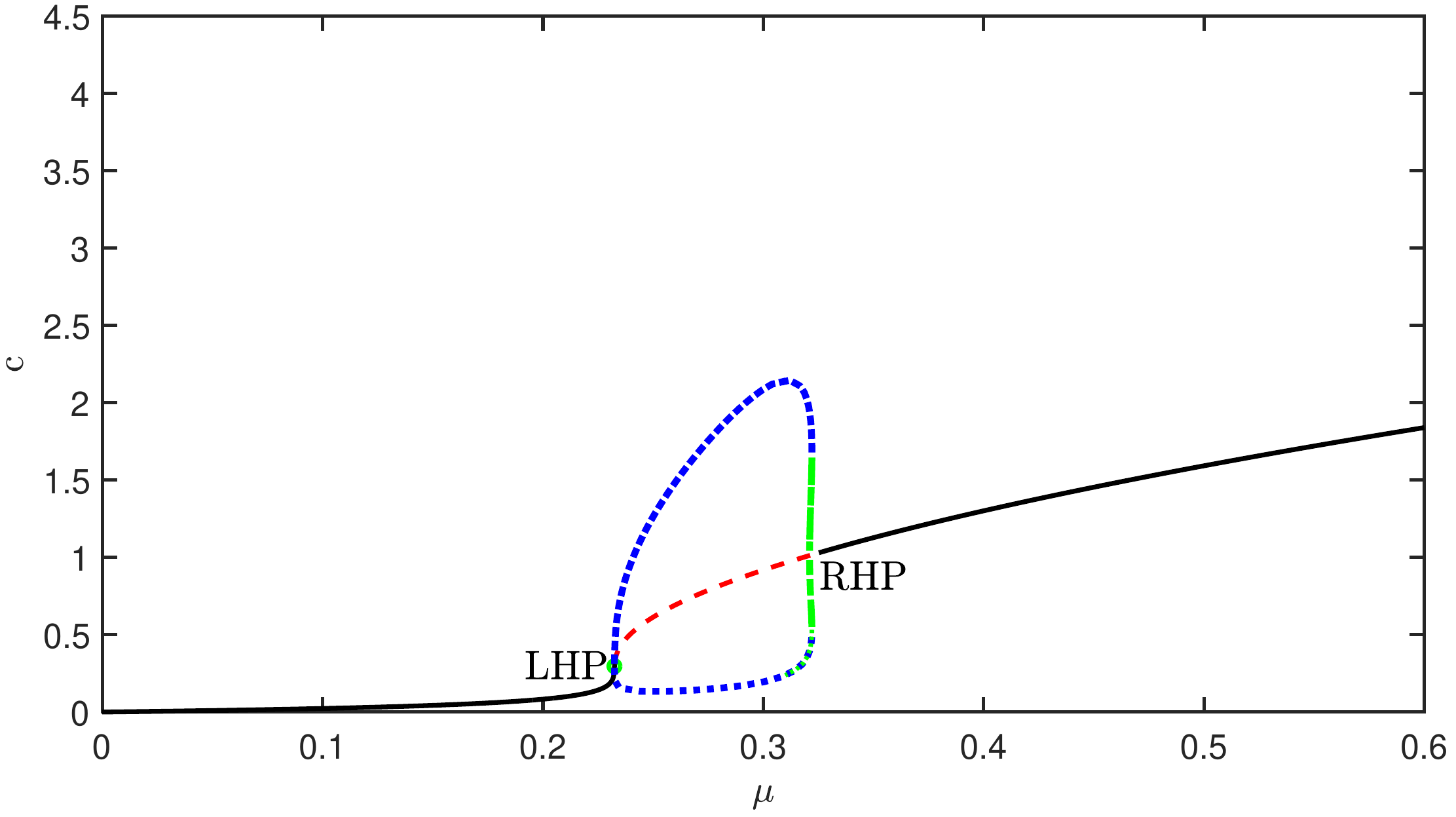}
		\caption{$\lambda=1$}\label{fig:AtriMechsLam1} 	
	\end{subfigure}
	\quad
	\begin{subfigure}[t]{1\textwidth}
		\centering
		\includegraphics[width=1\linewidth]{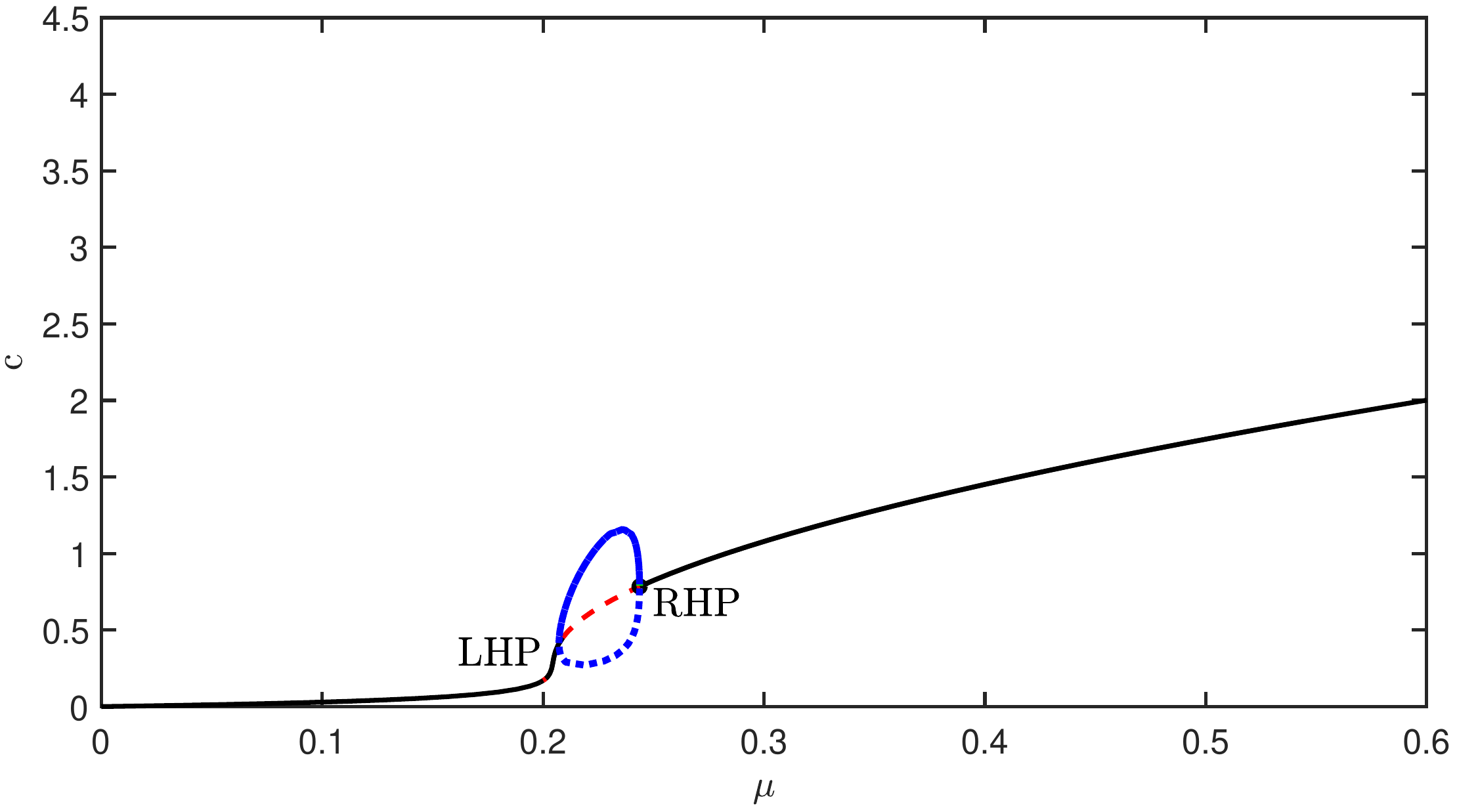}
		\caption{$\lambda=1.5$}\label{fig:AtriMechsLam1Pt5}
	\end{subfigure}
	\caption{Amplitude of calcium oscillations for the system \eqref{eq:Atri3Da-ND}--\eqref{eq:Atri3Dc-ND} when $\hat T(c)=\frac{10 c}{1+10 c}$, as $\mu$ is increased, for selected values of $\lambda$ (computed with XPPAUT and exported to Matlab for plotting). The LHP and the RHP are indicated. The stable limit cycles are represented by dots and the unstable limit cycles by the dash-dotted parts (respectively with blue and green colour online): (a) $\lambda=0.5$ (b) $\lambda=1$ (c) $\lambda=1.5$. As $\lambda$ increases, for any fixed $\mu$ the amplitude decreases until it becomes zero.}
\label{fig:AtriMechs1}
\end{figure}

In Figure \ref{fig:AtriMechs1} we plot the oscillation amplitude as a function of $\mu$, for three selected values of 
$\lambda$, using XPPAUT.  We see that as $\lambda$ increases the amplitude decreases until the oscillations vanish close to $\lambda=\lambda_{\rm  max}=1.69$, which agrees with the Hopf curve in Figure \ref{fig:HopfCurveFold}. We also observe that for $\lambda= 0.5$ and $1$, in Figures \ref{fig:AtriMechsLam0Pt5} and \ref{fig:AtriMechsLam1} respectively,  there are both stable and unstable limit cycles, and the right Hopf point is subcritical. Also, as $\lambda$ increases, the $\mu$-range of unstable limit cycles decreases until it vanishes; for $\lambda=1.5$ (Figure \ref{fig:AtriMechsLam1Pt5}) there are only stable limit cycles, and the right Hopf point has become supercritical. We see that as in the Atri model, the oscillation amplitude changes quite rapidly with $\mu$ in the mechanochemical system. 

\begin{center}
\begin{figure}
\includegraphics[width=1\textwidth]{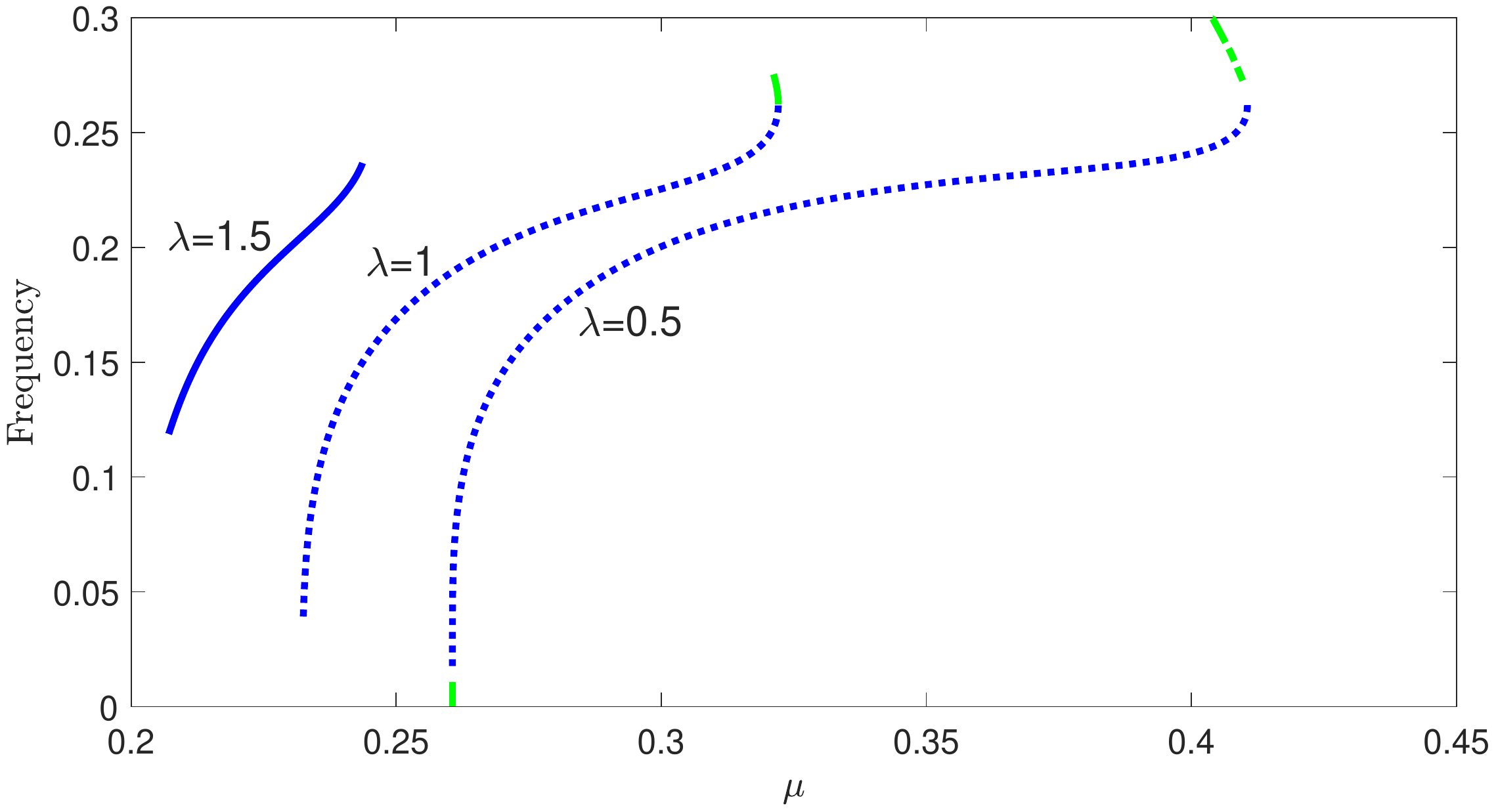}
\caption{Frequency of calcium oscillations for the system \eqref{eq:Atri3Da-ND}--\eqref{eq:Atri3Dc-ND} when  $\hat T(c)=\frac{10 c}{1+10 c}$ as a function of $\mu$ and for $\lambda=0.5, 1, 1.5$ (computed with XPPAUT and exported to Matlab for plotting). The stable limit cycles are represented by dots and the unstable limit cycles by dash-dotted parts (respectively with blue and green colour online).}
\label{fig:AtriMechsFreq}
\end{figure}
\end{center}

 In Figure \ref{fig:AtriMechsFreq} we plot the frequency of the limit cycles  as $\mu$ increases, for three values of $\lambda$, using XPPAUT. For $\lambda=0.5$ and  $\lambda=1$, the frequency increases  rapidly close to the LHP and the RHP and there is an `intermediate' region where the frequency varies slowly with $\mu$, as in the Atri system (see Figure \ref{fig:AtriBifnFrequency}). The `intermediate' region becomes smaller as $\lambda$ increases, and for $\lambda=1.5$  this region vanishes. We see that as $\lambda$ increases the frequency of oscillations decreases overall. 

Summarising, for any value of $\mu$ and $\lambda$ we can determine the range for oscillations using the parametric expressions \eqref{eq:MUvsC} and \eqref{eq:LAMBDAvsC}, and then use XPPAUT \cite{ermentrout2002simulating} or other continuation software to obtain their amplitude and frequency.

\begin{figure}[!ht] 
  \begin{subfigure}[b]{1\textwidth}
    \centering
    \includegraphics[width=0.65\linewidth]{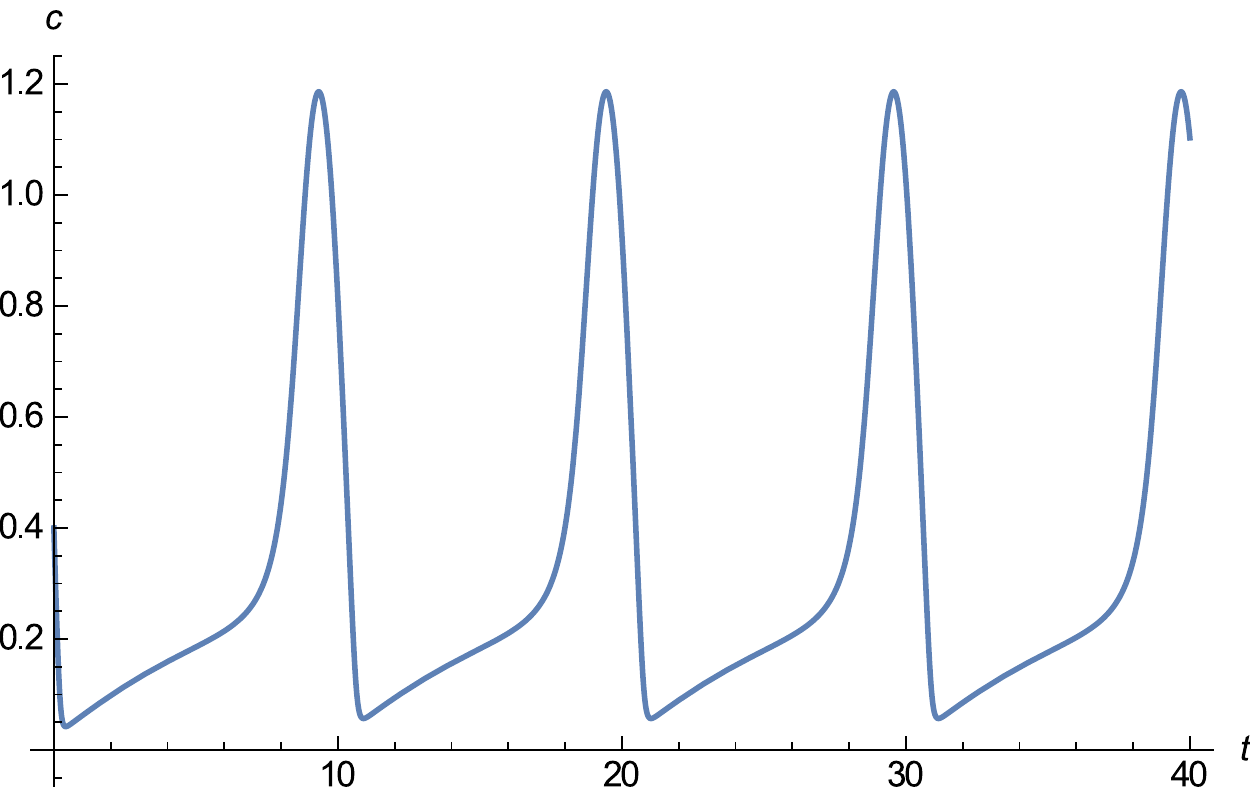}
    \caption{$\lambda=0$} 
  \label{fig:1a}		
  \end{subfigure}
\quad
  \begin{subfigure}[b]{1\textwidth}
    \centering
    \includegraphics[width=0.65\linewidth]{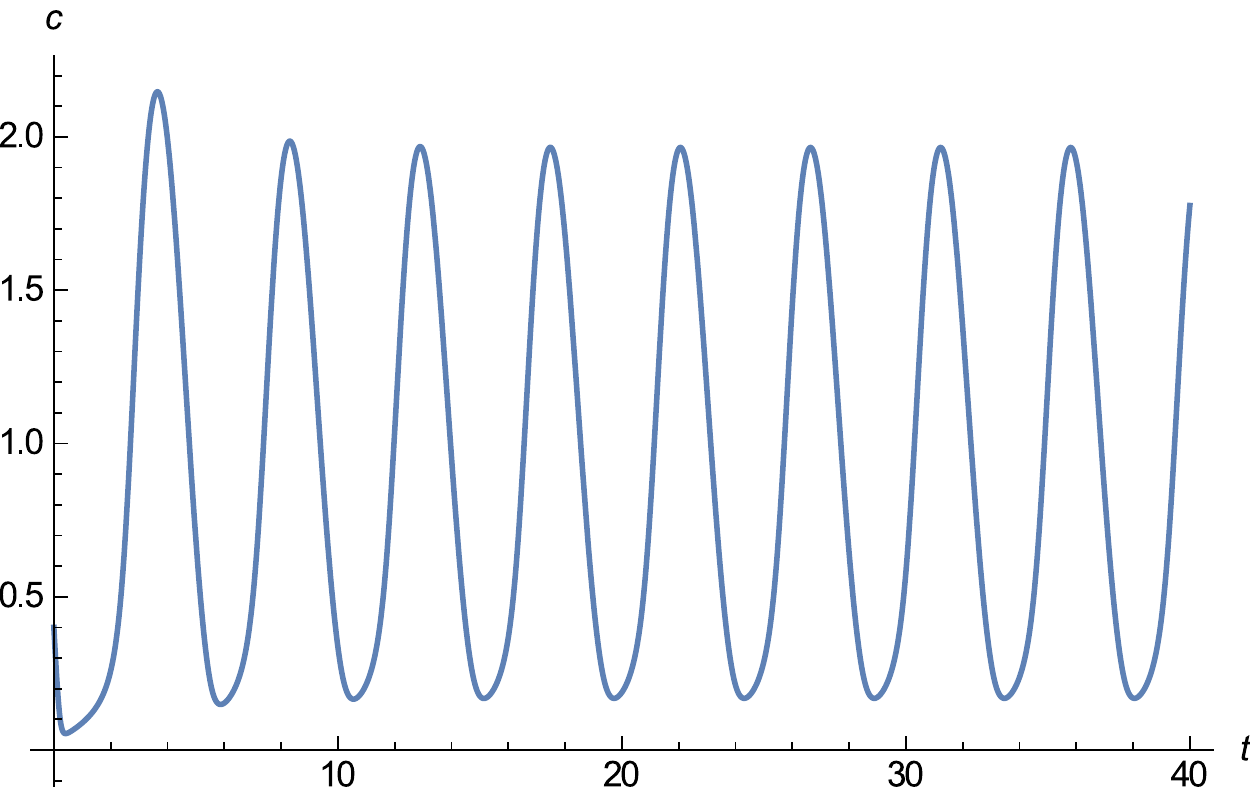}
		\caption{$\lambda=1$}\label{fig:1b}	
  \end{subfigure} 
\quad  
\begin{subfigure}[b]{1\textwidth}
    \centering
    \includegraphics[width=0.65\linewidth]{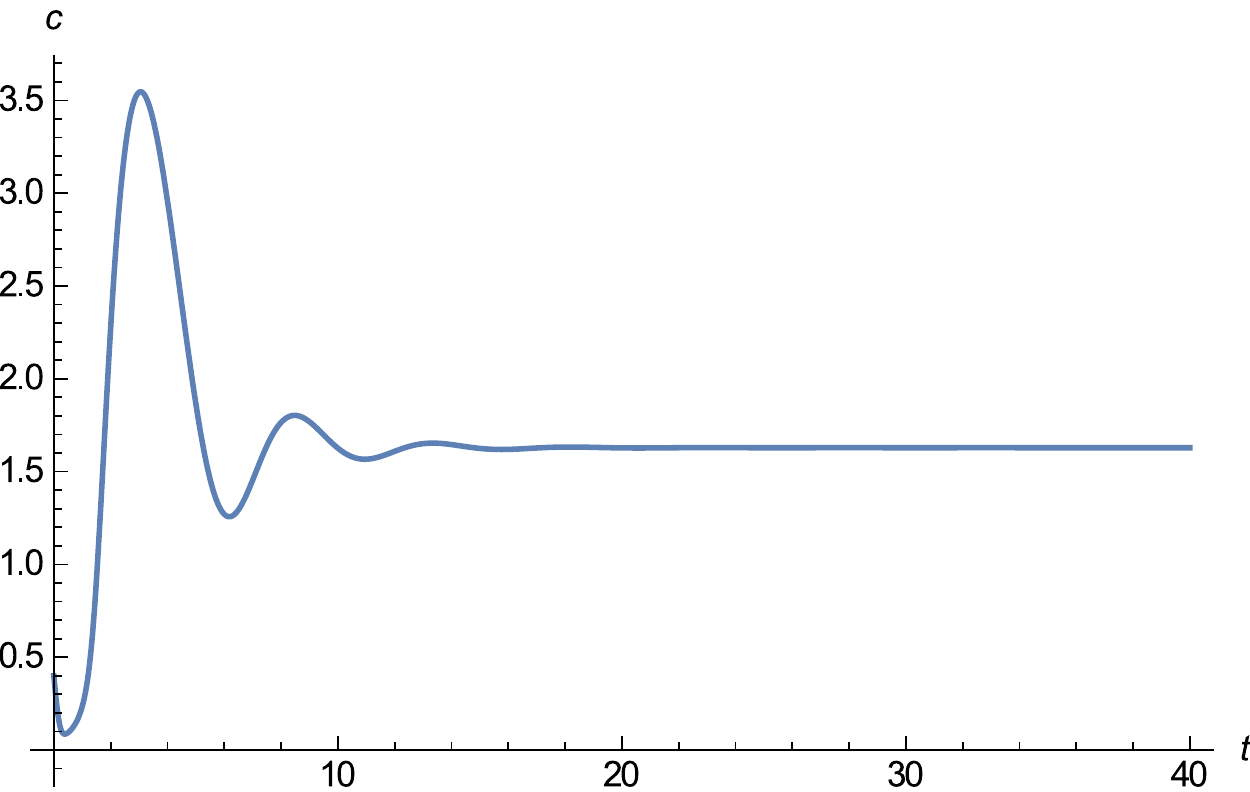}
		\caption{$\lambda=3$}\label{fig:1c}
  \end{subfigure}
  \caption{Evolution of $c(t)$ with time, solving the system \eqref{eq:Atri3Da-ND}--\eqref{eq:Atri3Dc-ND} numerically, when $\hat T(c)=\frac{10 c}{1+10 c}$, $\mu=0.289$ (a) $\lambda=0$ (Atri model): limit cycles (b) $\lambda=1$:  limit cycles with increased frequency and amplitude (c) $\lambda=3$: decaying solution; limit cycles (oscillations) disappear.}
\label{fig:1}
\end{figure}

In Figure \ref{fig:1} we plot the evolution of $c(t)$, solving \eqref{eq:Atri3Da-ND}--\eqref{eq:Atri3Dc-ND} numerically, for $\mu=0.3$ and selected values of $\lambda$; as expected from the bifurcation diagrams, the oscillations  are suppressed when $\lambda$ is sufficiently increased.

\subsubsection{Varying the cytosolic mechanical responsiveness factor}
\label{sec:mechResponse}
\begin{figure}	
	\centering
	\begin{subfigure}[t]{1\textwidth}
		\centering
		\includegraphics[width=1\linewidth]{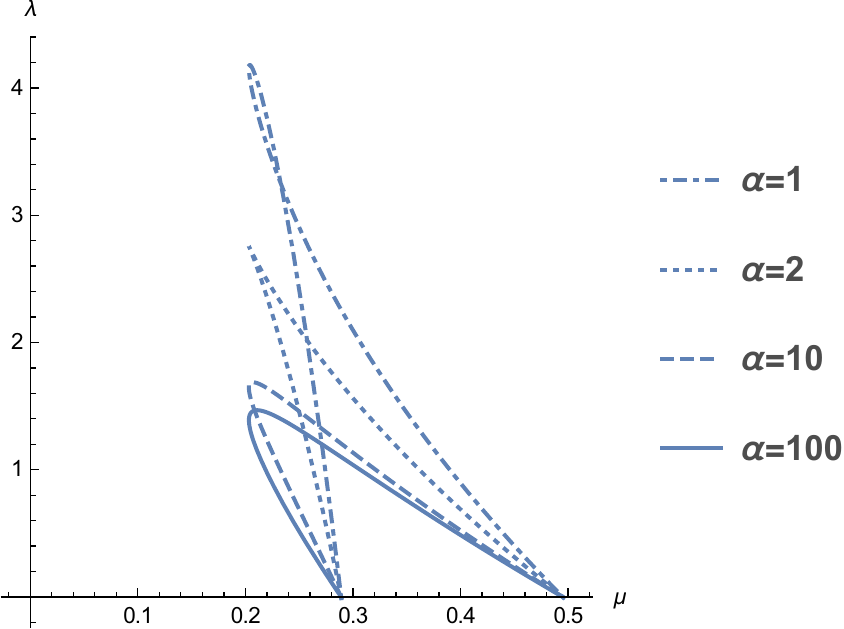}
		\caption{}\label{fig:HopfCurvesAlphaVaries}
	\end{subfigure}
	\quad
	\begin{subfigure}[t]{1\textwidth}
		\centering
		\includegraphics[width=1\linewidth]{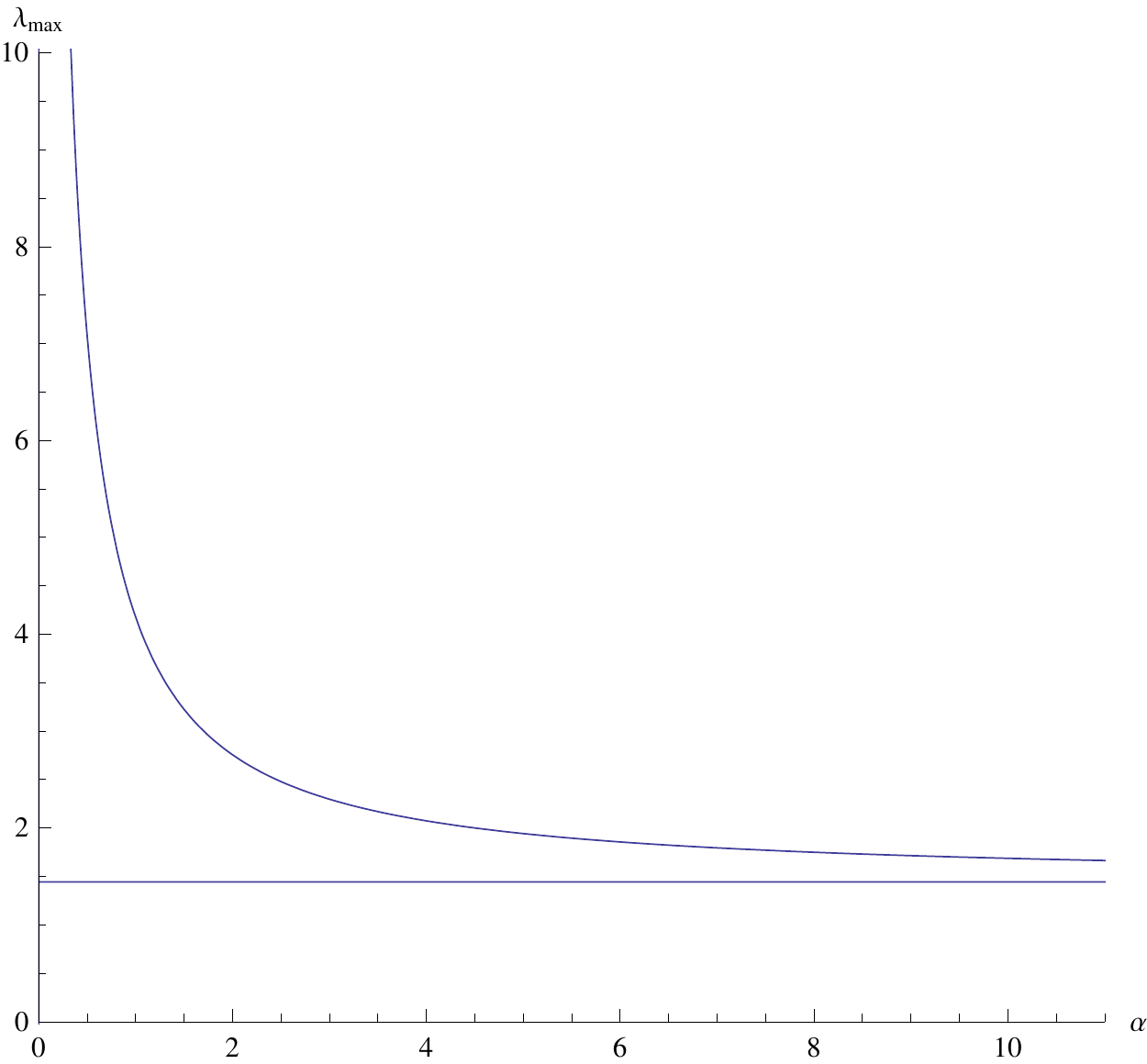}
		\caption{}\label{fig:pLamCritvsAlpha}
	\end{subfigure}
	\caption{(a) Hopf curves for the system \eqref{eq:Atri3Da-ND}--\eqref{eq:Atri3Dc-ND} and $\hat T(c)=\frac{\alpha c}{1+\alpha c}$, $\alpha=1, 2, 10, 100$ (see legend) (b) The maximum value of $\lambda$ for which oscillations are sustained, $\lambda_{\rm max}$, decreases with $\alpha$. Both plots are drawn using the parametric expressions \eqref{eq:MUvsC}--\eqref{eq:LAMBDAvsC}, in Mathematica.}\label{fig:}
\end{figure}

We now investigate if the Hopf curve changes qualitatively as the cytosol's mechanical responsiveness factor, $\alpha$, varies. In Figure \ref{fig:HopfCurvesAlphaVaries}, using the parametric expressions \eqref{eq:MUvsC}--\eqref{eq:LAMBDAvsC} we plot the Hopf curves for increasing values of $\alpha=1,2,10, 100$. We observe that the Hopf curve changes qualitatively; for $\alpha \approx 2$ it develops a cusp and for smaller values of $\alpha$ there is a ``bow-tie''. This geometrical change corresponds to yet another bifurcation, with $\alpha$ as a bifurcation parameter\footnote{The cusped Hopf curve is the bifurcation curve of a cusp catastrophe surface, according to the catastrophe theory developed by Zeeman \cite{zeeman1977}, and subsequently by Stewart and collaborators \cite{poston2014, stewart2014}.}. However, as for $\alpha=10$, oscillations always vanish for a sufficiently large value of $\lambda$,  $\lambda_{\rm max}$. 

We also observe that as $\alpha$ increases, $\lambda_{\rm max}$, the critical stretch activation value beyond which oscillations vanishes, decreases, i.e. oscillations are sustained for a smaller range of $\lambda$ values. To investigate this more systematically we have determined parametric expressions for $\lambda_{\rm \max}$ and $\alpha$ as functions of $c$, and we plot $\lambda_{\rm \max}(\alpha)$ in Figure \ref{fig:pLamCritvsAlpha}. We see that as $\alpha$ increases, $\lambda_{\rm max}$ decreases monotonically, and hence oscillations are sustained for an increasingly smaller range of $\lambda$, which agrees with Figure \ref{fig:HopfCurvesAlphaVaries}. Also, since 
$\lambda_{\rm max}(\alpha)$ asymptotes to a positive value as $\alpha \to \infty$, for any $\hat T(c)=\frac{\alpha c}{1+\alpha c}$, the system will always sustain some oscillations, irrespectively of the value of $\alpha$.  Therefore, we predict that for cytosols that are more responsive to calcium (higher $\alpha$), oscillations vanish at a lower $\lambda_{\rm \max}$.To test this experimentally the responsiveness of the cytosol to calcium should be manipulated whilst monitoring whether oscillations appear. The contractility of the cytosol could be manipulated by inhibiting Myosin II contractility using the ROCK inhibitor (Y-27632). 



However, since \eqref{eq:MUvsC} does not depend on $\hat T(c)$,  $\mu_{\rm min}$ is \emph{constant} and not zero for any $\alpha$. Therefore, as we expect, $IP_3$ is always required in order to obtain oscillations, for any $\lambda$  and any $\alpha$ but the minimum level of $IP_3$ does not change with $\alpha$. Also, for fixed $\lambda$, as $\alpha$, the mechanical responsiveness factor of the cytosol, increases, the $IP_3$ level required to induce oscillations also decreases. Additionally, for fixed $\mu$, as $\alpha$ increases $\lambda_{\rm max}$ reduces. 

Summarising, we conclude that as the cytosol's mechanical responsiveness increases a lower level of stretch activation is sufficient to sustain oscillations. Also, there will always be oscillations for some values of $\mu$ and $\lambda$ when the contraction stress is modelled as a Hill function of order 1.

\subsection{Hopf curves for $\hat T(c)$  a Hill function of order 2}
\label{sec:2ndHillFun}
\begin{figure}
\begin{center}
\includegraphics[width=0.85\textwidth]{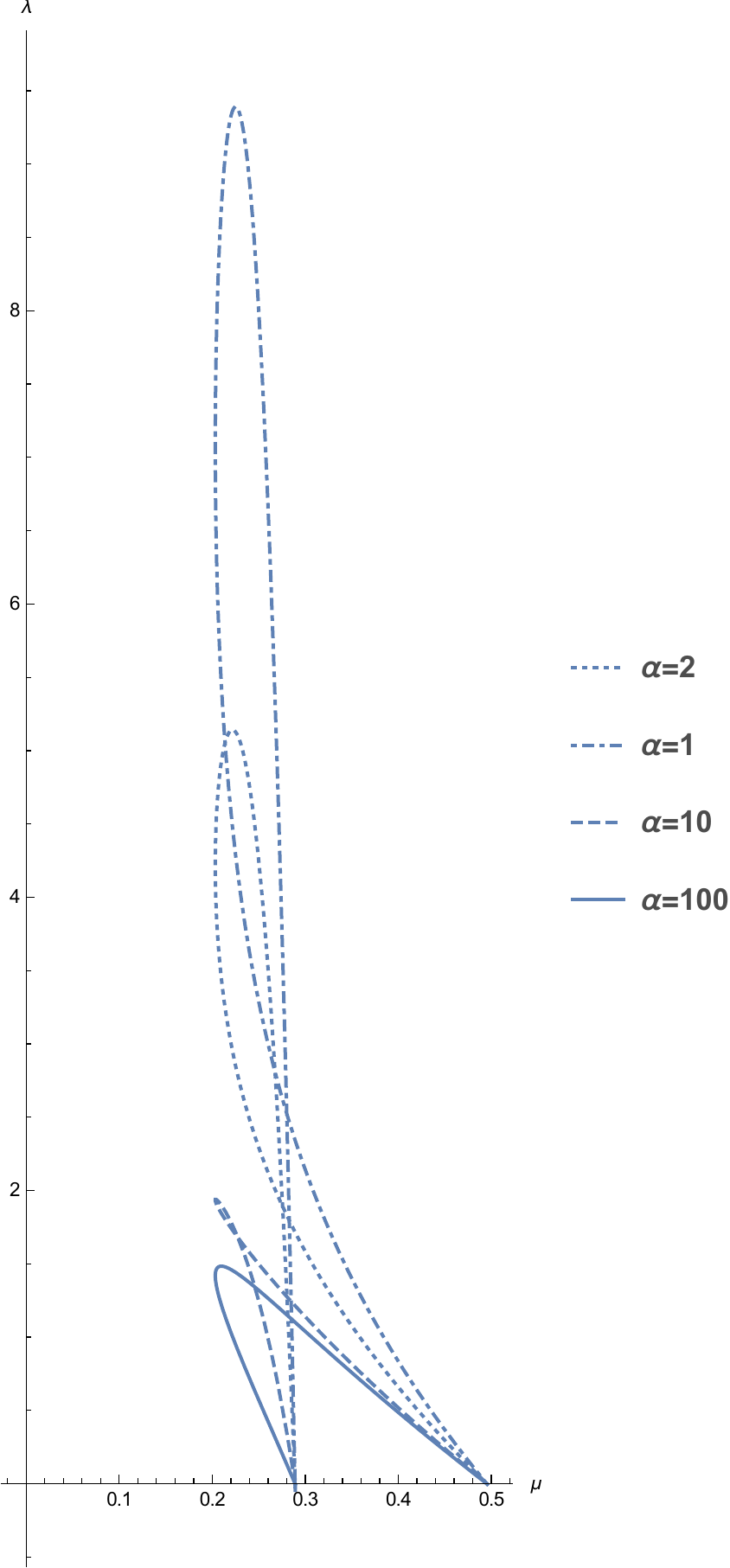}
 \caption{Hopf curves for the system \eqref{eq:Atri3Da-ND}--\eqref{eq:Atri3Db-ND} when $\hat T(c)=\frac{\alpha c^2}{1+\alpha c^2}$, for $\alpha=1, 2, 10, 100$ (see legend), drawn using the parametric expressions \eqref{eq:MUvsC}--\eqref{eq:LAMBDAvsC}, in Mathematica.}
\label{fig:HopfCurvesAlphaVariesNewHillFn}
\end{center}
\end{figure}

It is instructive to investigate whether a different functional form of $\hat T$ will change our conclusions. We thus model $\hat T(c)$  as a  Hill function of order 2, $\hat T(c)=\frac{\alpha c^2}{1+\alpha c^2}$, which models a cytosol which is less sensitive to calcium for low calcium levels than  $\hat T(c)=\frac{\alpha c}{1+\alpha c}$ but which saturates faster. 
In Figure \ref{fig:HopfCurvesAlphaVariesNewHillFn}  we plot the Hopf curves of the system \eqref{eq:Atri3Da-ND}--\eqref{eq:Atri3Db-ND} for increasing $\alpha$, the cytosolic mechanical responsiveness factor, using again the parametric expressions \eqref{eq:MUvsC}--\eqref{eq:LAMBDAvsC}.
Comparing  Figure \ref{fig:HopfCurvesAlphaVariesNewHillFn} with Figure \ref{fig:HopfCurvesAlphaVaries} we see that the Hopf curves have the same qualitative behaviour for the two Hill functions. Oscillations can be sustained for any value of $\alpha$ and they always vanish for a sufficiently large value of $\lambda$, Also, as in the Hill function of order 1 as $\alpha$ increases $\lambda_{\rm max}$ decreases while $\mu_{\rm min}$ is constant. Also, a cusp again develops but for the Hill function of order 2 the value of $\alpha$ where this occurs increases. We conclude that the conclusions are robust to the change of the Hill function. In future work Hill functions of higher order or other functional forms of $T$ can be investigated.

\newpage
\section{Summary, conclusions and future research directions}
\label{sec:Conclusions}
A wealth of experimental evidence has accumulated which shows that  many types of cells release calcium in response to mechanical stimuli but also that calcium release causes cells to contract. Therefore, studying this mechanochemical coupling is important for elucidating a wide range of body processes and diseases. In this work we have focused attention on embryogenesis, where the interplay of calcium and mechanics is shown to be essential in Apical Constriction (AC), an essential morphogenetic process which, if disrupted, leads to embryo abnormalities \cite{christodoulou2015cell}. 

We have presented a new analysis of experimental data that supports the existence of a two-way mechanochemical coupling between calcium signalling and contractions in embryonic epithelial cells involved in Apical Constriction.

We  have then developed a simple mechanochemical ODE model that consists of an ODE for $\theta$, the cell apical dilation, derived consistently from a full, linear viscoelastic \emph{ansatz} for a Kelvin-Voigt material, and two ODEs governing, respectively, the evolution of calcium and the proportion of active $IP_3$ receptors. The two latter ODEs are based on the well-known, experimentally verified, Atri model for calcium dynamics \cite{atri1993single}. An important feature of our model is the \emph{two-way} coupling between calcium and mechanics which was proposed for the first time in models by Oster, Murray and collaborators \cite{murray1984generation, oster1984mechanochemistry, murray1988, murray2001}. However, in those models hypothetical bistable calcium dynamics were considered whereas here we have updated those models with recent advances in calcium signalling, as encapsulated by the Atri model.  We have also modelled the calcium-dependent contraction  stress in the cytosol as a Hill function $\hat T(c)$, since experiments indicate that the mechanical responsiveness of the cytosol to calcium saturates for high calcium levels. 

The early mechanochemical models included an ad hoc stretch activation calcium flux, $\lambda \theta$,  in the calcium ODE. Here, we have also derived, for the first time, this ``stretch-activation" flux as a ``bottom-up" contribution from stretch sensitive calcium channels (SSCCs), thus expressing the parameter $\lambda$ as a combination of the structural characteristics of a SSC; $\lambda$ can also be thought of as a coupling parameter between calcium signalling and mechanics. Despite an extensive literature search we could not find experimental measurements for SSCCs; this could be a direction for future experiments. 

For \emph{any} $\hat T(c)$, we have analytically identified the parameter regime in the $\mu$--$\lambda$ plane corresponding to calcium oscillations and applied this result in two illustrative examples,  $\hat T(c)=\alpha c/(1+\alpha c)$ and $\hat T(c)=\alpha c^2/(1+\alpha c^2)$. In both cases, as $\lambda$ increases, the oscillations are eventually suppressed at a critical $\lambda$,  $\lambda_{\rm \max}$---see, respectively, Figures \ref{fig:HopfCurvesAlphaVaries} and \ref{fig:HopfCurvesAlphaVariesNewHillFn}. The prediction is in agreement with experiments \cite{christodoulou2015cell} where a high, non-oscillatory calcium state is associated with a very high stress in the cytosol and continuous contraction (Figure 5D). This high-calcium, high-stress state is associated with failure of Apical Constriction and consequently with defective tissue morphogenesis. This makes sense since calcium oscillations are the `information carrier' in cells so we indeed expect that if they vanish the task at hand, in this case AC, will not be performed correctly. In summary, we have shown that there are scenarios where mechanical effects significantly affect calcium signalling and this is a key result of this work.

For $\hat T(c)=\alpha c/(1+\alpha c)$ we have also shown analytically that as $\alpha$, the mechanical responsiveness factor of the cytosol increases,  $\lambda_{\rm \max}$ decreases but it never becomes zero (see Figure \ref{fig:pLamCritvsAlpha}). This means that for any $\alpha$, there will always be a $\mu$-$\lambda$ region for which oscillations are sustained.  Furthermore, for the illustrative example of $\hat T(c)=10 c/(1+10 c)$ we have determined numerically the oscillation amplitude and frequency  as the bifurcation parameters $\mu$ and $\lambda$ vary, using XPPAUT. We found that the behaviour is qualitatively similar to the Atri model (see Figure \ref{fig:AtriBifnFrequency}) for lower $\lambda$ values but that it changes for larger $\lambda$ values  (see Figure \ref{fig:AtriMechsMu}). We found that, as $\lambda$ increases the amplitude of oscillations decreases (see Figure \ref{fig:AtriMechs1}) but their frequency decreases (see Figure \ref{fig:AtriMechsFreq}). More experiments could be undertaken to test these predictions. 

In the experiments of \cite{christodoulou2015cell} the calcium-induced stress saturates to a non-zero level as calcium levels increase but in other cell types it is possible that the cell can relax back to baseline stress and in such a case $\hat T(c)$ cannot be modelled as a Hill function. Experiments could be undertaken also in other calcium-induced mechanical processes to determine the appropriate form of $\hat T(c)$ and the model could then be modified appropriately.

Another approximation we have made is that the mechanical properties of the cell (Young's modulus, Poisson ratio, viscosity) are constant. However, their values can change significantly with space and also with embryo stage \cite{brodland2006cell, luby1999cytoarchitecture, zhou2009actomyosin}. One of the next steps in the modelling would be to take these variations into account. 

Due to the complexity of calcium signalling all models introduce approximations. One important approximation in this work is that we neglect stochastic effects, even though the opening and closing of $IP_3$ receptors and of the SSCCs is a stochastic process. However, the deterministic models still have good predictive power, whilst being more amenable to analytical calculations \cite{thul2014, cao2014}. A multitude of deterministic and stochastic calcium models have been developed \cite{atri1993single,wilkins1998,goldberg2010,gracheva2001stochastic,sneyd1994model,sneyd1998calcium,timofeeva2003}; see also the comprehensive reviews \cite{sneyd2003, thul2014, rudiger2014stochastic} and the books \cite{keener1998, dupont2016}, among others.  Future work could involve developing stochastic mechanochemical models.

The interplay of mechanics and calcium signalling in non-excitable cells is  important in processes occurring not only in embryogenesis but also in wound healing and cancer, amongst many others, and  more efforts should be devoted in developing appropriate mechanochemical calcium models that would help elucidate the currently many open questions. In this connection, the insights we have obtained from the simple model we have developed here are a first step in this direction. We will aim to extend to models in more realistic geometries. Moreover, we have fixed all parameters here, except $\mu$, $\lambda$ and $\alpha$; and the variation of other parameter values may lead to other bifurcations and biologically relevant behaviours. 

Finally, the newly discovered SSCCs merit much more experimental investigation and modelling; in this work we have adopted a simple model for their behaviour, assuming them quasisteady and also made restricting assumptions about their opening and closing rates. In further experimental work, the parameter values associated with SSCCs should be measured and perhaps more sophisticated models for SSCCs should be developed.

\begin{acknowledgements}
We thank James Sneyd, Vivien Kirk, Ruediger Thul,  Lance Davidson, Abdul Barakat, Thomas Woolley, Iina Korkka and Bard Ermentrout for valuable discussions. Katerina Kaouri also acknowledges support from two STSM grants awarded by the COST Action TD1409 (Mathematics for Industry Network, MI-NET) for research visits to Oxford University.
\end{acknowledgements}

\bibliographystyle{plain}
\bibliography{refsCW-Apr18}

\appendix
\section{Appendix}

\subsection{Supplementary information on the presented experimental results}
\label{sec:suppInfo}
Figure 1:
Plot of surface area (measured in $\mu m^2$) and calcium oscillation amplitude over time of a single embryonic, epithelial cell undergoing Apical Constriction in Xenopus \cite{christodoulou2015cell}. Measurements were taken every 10 seconds from a time lapse movie of a stage 9 Xenopus embryo expressing Lulu-GFP + GECO-RED (see \cite{christodoulou2015cell}). To normalise the surface area, all the area values were divided by the first measured value (which is the largest since the surface area is decreasing with time).   For measuring calcium oscillation amplitudes, the signal intensity of the non-ratiometric calcium sensor (GECO-RED) was also measured over time. For normalization, all values were divided by the highest signal intensity value. Note that we tracked the surface area and calcium level of cells induced to undergo AC at gastrula stage in order to decouple AC from other mophogenetic movements, like mediolateral junction shrinkage and convergent extension, which take place in later embryogenic stages and which would also influence the cell shape and surface area.  

Figure 2:
(a) Plot of surface area ($\mu m^2$) and calcium oscillation frequency over time from 10 neural plate cells of stage 16 Xenopus embryo using the mem-GFP +GECO RED sensor molecule. The average surface area of each cell was evaluated for four time intervals; 0-10, 10-20, 20-30 and 30-40 minutes. For normalization for each cell, the average surface area in each time period was divided by the average surface area of the first period (0-10 min)   The calcium oscillation frequency in each cell was calculated by counting the number of calcium oscillations, in each time interval. This value was then divided by 10 since there are 10 minutes in each time interval.
(b) Plot of average calcium oscillation amplitude of 10 neural plate cells (same cells as in (a)). The signal intensity of the non-ratiometric calcium sensor (GECO-RED) was measured per calcium oscillation in each of the cells over time. For normalization, the values were then divided by the highest intensity value. The average value in each of the four time intervals was plotted. 

\subsection{Analysis of the Atri model}
\label{sec:LinStabDetails}
\subsubsection{Linear stability analysis}
\begin{figure}
\begin{center}
  \includegraphics[width=0.65\textwidth]{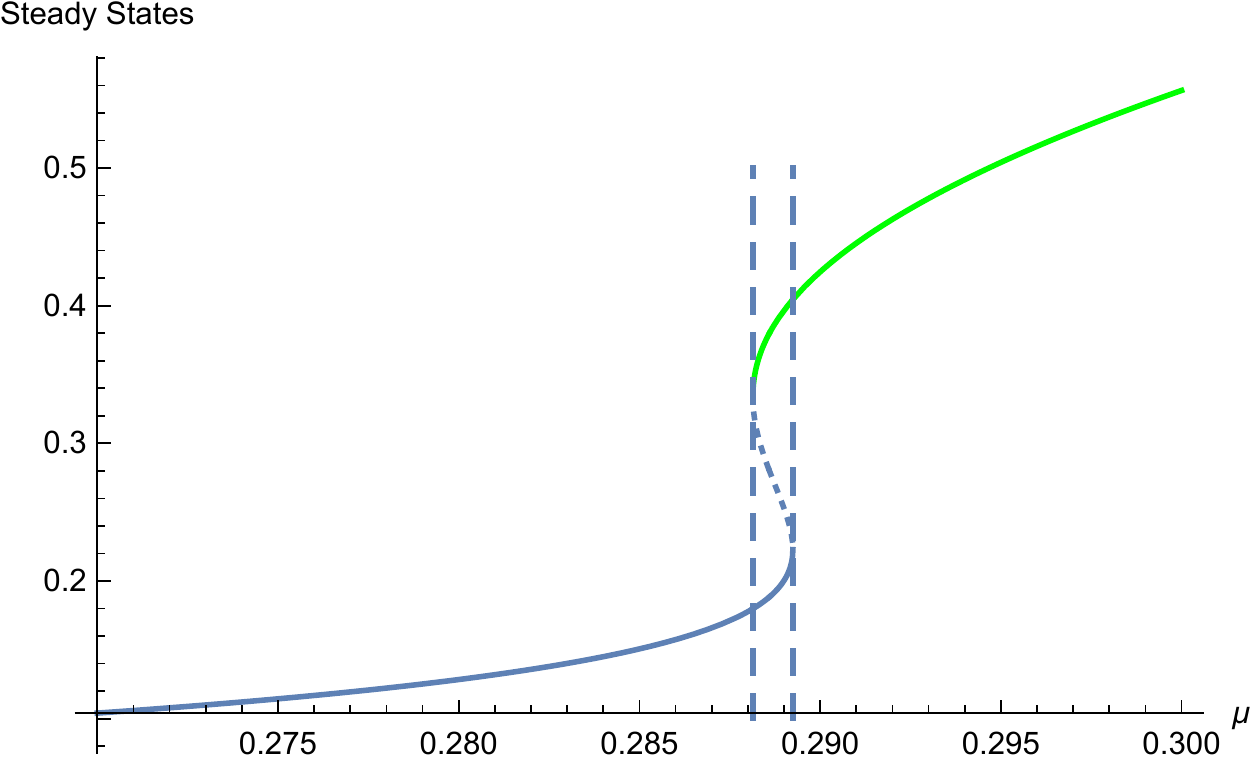}
    \caption{The steady states of \eqref{eq:Atri1}-\eqref{eq:Atri2} as a function of the bifurcation parameter $\mu$, as obtained using the analytical expression \eqref{eq:mu-SS}. As $\mu$ increases, from small to large there is one steady state, a double (degenerate) steady state at $\mu_1=0.28814$, then three steady states, then a double (degenerate) steady state at $\mu_2=0.28925$, and for values of $\mu$ larger than $0.28925$ one steady state.}
    \label{fig:pcFunctionmu}
\end{center} 
\end{figure}

The steady states (S.S.) of \eqref{eq:Atri1}-\eqref{eq:Atri2} are the intersections of the nullclines of the system. Setting
\begin{align}
\label{eq:Nullcline1}
F=0 \implies h&=\frac{\Gamma}{\mu K_1}\frac{c(1+c)}{(K+c)(b+c)},\\
\label{eq:Nullcline2}
G=0 \implies h&=\frac{1}{1+c^2}.
\end{align}
we obtain
\begin{align}
\label{eq:mu-SS}
\mu K_1\frac{1}{1+c^2}\frac{b+c}{1+c}-\frac{\Gamma c}{K+c}=0,
\end{align}
which can be cast as a quartic in $c$. (Note that \eqref{eq:mu-SS-3D} reduces to \eqref{eq:mu-SS} for $\lambda=0$, as expected.) In Figure  \ref{fig:pcFunctionmu} we plot the equilibrium curve \eqref{eq:mu-SS} in order to visualise the number of steady states and the corresponding value(s) of $c$, as $\mu$ is increased. 
The qualitative behaviour of the solutions of the system can be determined by plotting the nullclines \eqref{eq:Nullcline1} and  \eqref{eq:Nullcline2}. When the nullclines cross the system has a steady state, and when they touch the system has a double (degenerate) steady state. Nullcline \eqref{eq:Nullcline1} passes through the origin of the ($c$,$h$) plane, has a maximum at $h=h_M$ and saturates to the constant value $h=\frac{\Gamma}{\mu K_1}$ as $c \to \infty$. $h_M$ can be found analytically by differentiating \eqref{eq:Nullcline1}:
\begin{align}
\frac{dh}{dc}&=\frac{\Gamma}{\mu K_1}\frac{c^2(b+K-1)+2bcK+bK}{(K+c)^2(b+c)^2},
\label{eq:dhdc}
\end{align}
and setting $dh/dc=0$, which leads to a quadratic equation for $c$. Rearranging, and  since $c>0$, we discard the negative root, obtaining
\begin{align}
c_M(b,K)&=\frac{bK+\sqrt{(1-b)b(K-K^2)}}{1-b-K}\label{eq:cM}\\
\intertext{and, hence, substituting \eqref{eq:cM} in \eqref{eq:Nullcline1} we obtain}
h_M&=h(c_M)=\frac{\Gamma}{\mu K_1}\frac{c_M(1+c_M)}{(K+c_M)(b+c_M)}.
\end{align}
Therefore, $h_M$ scales with $\Gamma/(\mu K_1)$ and depends on the parameters $K$ and $b$ in a much more complicated manner. For the parameter values from \cite{atri1993single} we have $c_M=0.169$ and $h_M=0.279/\mu$.

Nullcline \eqref{eq:Nullcline2} is a decreasing function of $c$; it has a maximum at (0,1) and tends to $0$ as $c \to \infty$. For $\mu_1=0.28814$ and $\mu_2=0.28925$ the nullclines touch and there is a double steady state; for values of  $\mu<\mu_1$  and  $\mu>\mu_2$ there is one S.S. and there are three S.S. for $\mu_1<\mu <\mu_2$. ($\mu_1$ and $\mu_2$ are obtained by differentiating \eqref{eq:mu-SS} and finding the roots of $\frac{d\mu}{dc}=0$.) Note that we present the values of $\mu$ with five decimal places because the bifurcation analysis depends sensitively on $\mu$, as we will see later.

\captionsetup[subfigure]{textfont=normalfont,singlelinecheck=on,justification=centering} 
We then linearise the system near the steady states. 
We determine the Trace (Tr), Determinant (Det) and Discriminant (Discr) of the Jacobian of the system as follows
\begin{align*}
&{\rm Tr}=F_{c}+G_h=F_c-1=\frac{\Gamma}{(K+c)}\left(-\frac{K}{K+c}+\frac{(1-b)\Gamma c}{(1+c)(b+c)}\right)-1\\
&{\rm Det}=F_cG_h-F_hG_c=-\frac{\Gamma}{(K+c)}\left(-\frac{K}{K+c}+\frac{(1-b)\Gamma c}{(1+c)(b+c)}\right)+\frac{2\Gamma c^2}{(1+c^2)(K+c)}\\
&{\rm Discr}=({\rm Tr})^2-{\rm 4Det}.
\end{align*}

Taking the parameter values of \cite{atri1993single} we identify the bifurcations of the system as $\mu$ increases by determining where the Tr, Det, and Discr change sign. We find a richer bifucation structure as $\mu$ increases. This behaviour was not analysed in such detail in \cite{atri1993single} or in later literature. Given the very sensitive dependence on precise values of $\mu$, these details are probably of more mathematical interest than of biological significance. The parameter values are summarised in Table \ref{tab:Table1}.
\begin{itemize} 
\item[$\bullet$] $0<\mu<0.27828$: one stable node.
\item[$\bullet$] $\mu=0.27828$: the stable node becomes a stable spiral (bifurcation Discr=0)
\item[$\bullet$] $\mu=0.28814$: Stable spiral present. Also, a saddle and an unstable node (UN) emerge (bifurcation Det=0, fold point)
\item[$\bullet$] $\mu=0.28900$: the stable spiral becomes an unstable spiral. The other two S.S. are still a saddle and an unstable node. (Tr=0, Hopf bifurcation) 
\item[$\bullet$]  $\mu=0.28924$ the unstable spiral becomes an unstable node, and we have two unstable nodes and a saddle (Discr=0)
\item[$\bullet$] $\mu=0.28925$: one unstable node (Det=0, fold point)
\item[$\bullet$] $\mu=0.28950$: the unstable node becomes an unstable spiral (Discr=0)
\item[$\bullet$] $\mu=0.49500$: the unstable spiral becomes a stable spiral. (Tr=0, Hopf bifurcation)
\end{itemize}
From the regimes identified above we are particularly interested in the regime of relaxation oscillations, since their amplitude and/or frequency encodes the information in calcium signals. Relaxation oscillations are sustained for $0.28900 \leq \mu \leq 0.49500$ since at $\mu=0.28900$ a Hopf bifurcation (HB) arises, the stable spiral becomes unstable, and we expect relaxation oscillations (limit cycles) in the nonlinear system. As $\mu$ increases the unstable spiral becomes a stable spiral close to $\mu=0.495000$ and the limit cycles eventually vanish.

\begin{table}
\centering
\begin{tabular}{ |c|c|  }
\hline
 \multicolumn{2}{|c|}{Parameter values} \\
 \hline
     Parameter & Value \\
 \hline
 b   & 0.111   \\
 $k_1$&   $0.7\mu M$  \\ 
 $k_f$ &  $16.2\mu M/s$ \\
$k_\mu$&  $0.7\mu M$  \\
$\gamma$    &$2\mu M/s$ \\
 $k_{\gamma}$& $0.1\mu M$    \\
$\beta$ &  $0-0.02 \mu M/s$   \\
 $k_2$ & $0.7\mu M$  \\
 $\tau_h$&  $2s$  \\
 \hline
\end{tabular}
\caption{Parameter values, taken from \cite{atri1993single}}
\label{tab:Table1}
\end{table}

\subsection{Linear stability analysis of the mechanochemical model with no $IP_3$}
\label{sec:muZero}
Here we analyse the case $\mu=0$. Biologically, this corresponds to treating the cell with thapsigargin so that all calcium is depleted from the ER and there is no flux from the ER. The model \eqref{eq:Atri3Da-ND}-\eqref{eq:Atri3Db-ND} simplifies to
\begin{align}
\label{eq:Eqc-mu0}
\frac{dc}{dt}&=-\frac{\Gamma c}{K+c}+\lambda \theta,\\
\label{eq:Eqtheta-mu0}
\frac{d\theta}{dt}&=-\theta+\hat T(c),\\
\label{eq:Eqh-mu0}
\frac{dh}{dt}&=\frac{1}{1+c^2}-h.
\end{align}
Equation \eqref{eq:Eqh-mu0} is decoupled from equations \eqref{eq:Eqc-mu0} and \eqref{eq:Eqtheta-mu0}, and we can thus continue with a two-dimensional analysis for equations \eqref{eq:Eqc-mu0}--\eqref{eq:Eqtheta-mu0}. The steady states satisfy $\displaystyle{\lambda \hat T(c)=\frac{\Gamma c}{K+c}}$.
The Jacobian of the system \eqref{eq:Eqc-mu0}--\eqref{eq:Eqtheta-mu0} has entries
\begin{align*}
M_{11}=-\frac{\Gamma K}{(K+c^{\star})^2},\,\,M_{12}=\lambda,\,\,M_{21}=T'(c^{\star}),\,\,M_{22}=-1.
\end{align*}
Hence
\begin{align*}
{\rm Tr}(M_1)=-\frac{\Gamma K}{(K+c^{\star})^2}-1<0&,\,\,{\rm Det}(M_1)=\frac{\Gamma K}{(K+c^{\star})^2}-\lambda \hat T'(c^{\star})\\
{\rm Discr}(M_1)=&\left(\frac{\Gamma K}{(K+c^{\star})^2}-1 \right)^2+4\lambda \hat T'(c^{\star})>0.
\end{align*} 
Therefore,  for any $\hat T(c)$, ${\rm Discr}(M_1)>0$, ${\rm Tr}(M_1)<0$ always, and ${\rm Det}(M_1)$ can be negative or positive, the steady states can only be stable nodes or saddles, and oscillations cannot be sustained. 
For some choices of $\hat T$ a non-zero steady state may exist and this means biologically that even without calcium flux from the ER a non-zero calcium concentration can be sustained in the cytosol due to the stress-induced calcium release.

For $\hat T(c)$ as given in \eqref{eq:Tmodel}, there is always a steady state ($c^{\star},\theta^{\star}$)=($0,0$). A second steady state 
\begin{align}
\label{eq:SS2}
c^{\star}=\frac{\delta-K}{1-\alpha \delta},\,\,\textrm{ where } \delta=\frac{\Gamma}{\alpha \lambda},
\end{align}
exists if
\begin{align}
\label{eq:SS2ConditionsA}
&\delta>K\,\, \textrm{and}\,\, \alpha \delta<1\,\Rightarrow \Gamma  < \lambda<\frac{\Gamma}{\alpha K}\\
\label{eq:SS2ConditionsB}
\textrm{ or }\,\,&\delta< K\,\,\textrm{ and }\,\,\alpha \delta>1\,\Rightarrow \frac{\Gamma}{\alpha K} < \lambda< \Gamma.
\end{align}
We can easily show that the steady state (0,0) loses its stability and becomes a saddle when the second stable S.S. emerges (as a stable node). This means that there is a range of $\lambda$ values for which the system sustains a non-zero calcium concentration even without a CICR flux.
For even larger values of $\lambda$ there is no steady state and the model ceases to be biologically relevant.
\end{document}